\documentclass[twocolumn,english,aps,prx,showpacs,superscriptaddress,nofootinbib,longbibliography]{revtex4-1}
\usepackage{multirow}
\usepackage{bbm}
\usepackage{mathrsfs}
\usepackage{epsfig}
\usepackage{graphicx}
\usepackage{amsfonts}
\usepackage[figuresright]{rotating}
\usepackage{amssymb}
\usepackage{amsmath}
\usepackage{dcolumn}
\usepackage{bm}
\usepackage{color}
\usepackage{psfrag}
\usepackage{natbib}
\usepackage{graphicx,amsfonts,amssymb,amsmath}
\usepackage{overpic}
\usepackage{algpseudocode} 
\usepackage{algorithm}

\usepackage[T1]{fontenc}
\usepackage[latin9]{inputenc}
\usepackage{babel}
\usepackage{float}
\usepackage{booktabs}
\usepackage[unicode=true,
 bookmarks=false,
 breaklinks=false,pdfborder={0 0 1},backref=false,colorlinks=true]
 {hyperref}
\hypersetup{
 urlcolor=blue,citecolor=magenta}

\makeatletter

\providecommand{\tabularnewline}{\\}

\usepackage{babel}
\usepackage{braket}
\PassOptionsToPackage{normalem}{ulem}
\newcommand{\Eq}[1]{Eq.~(\ref{#1})}

\newcommand{\comments}[1]{}
\makeatother

\begin{document}

\title{Equivalence of restricted Boltzmann machines and tensor network states}

\author{Jing Chen}
\affiliation{Institute of Physics, Chinese Academy of Sciences, P.O. Box 603, Beijing
100190, China}
\affiliation{School of Physical Sciences, University of Chinese Academy of Sciences, Beijing, 100049, China}
\affiliation{Center for Computational Quantum Physics, Flatiron Institute, New York, 10010, USA}

\author{Song Cheng}
\affiliation{Institute of Physics, Chinese Academy of Sciences, P.O. Box 603, Beijing
100190, China}
\affiliation{School of Physical Sciences, University of Chinese Academy of Sciences, Beijing, 100049, China}

\author{Haidong Xie}
\affiliation{Institute of Physics, Chinese Academy of Sciences, P.O. Box 603, Beijing
100190, China}
\affiliation{School of Physical Sciences, University of Chinese Academy of Sciences, Beijing, 100049, China}

\author{Lei Wang}
\email{wanglei@iphy.ac.cn}
\affiliation{Institute of Physics, Chinese Academy of Sciences, P.O. Box 603, Beijing
100190, China}

\author{Tao Xiang}
\email{txiang@iphy.ac.cn}
\affiliation{Institute of Physics, Chinese Academy of Sciences, P.O. Box 603, Beijing
100190, China}
\affiliation{Collaborative Innovation Center of Quantum
Matter, Beijing 100190, China}

\begin{abstract}
  The restricted Boltzmann machine (RBM) is one of the fundamental building blocks of deep learning. RBM finds wide applications in dimensional reduction, feature extraction, and recommender systems via modeling the probability distributions of a variety of input data including natural images, speech signals, and customer ratings, etc. We build a bridge between RBM and tensor network states (TNS) widely used in quantum many-body physics research. We devise efficient algorithms to translate an RBM into the commonly used TNS. Conversely, we give sufficient and necessary conditions to determine whether a TNS can be transformed into an RBM of given architectures. Revealing these general and constructive connections can cross-fertilize  both deep learning and  quantum-many body physics. Notably, by exploiting the entanglement entropy bound of TNS, we can rigorously quantify the expressive power of RBM on complex datasets. Insights into TNS and its entanglement capacity can guide the design of more powerful deep learning architectures. On the other hand, RBM can represent quantum many-body states with fewer parameters compared to TNS, which may allow more efficient classical simulations.
\end{abstract}
\maketitle


\section{Introduction \label{sec:introduction}}
Deep learning is transforming the world with its far-reaching applications in computer vision, speech recognition, natural language processing, recommender systems, etc~\cite{LeCun:2015dt,Goodfellow-et-al-2016-Book}.
At the core of many of these applications are the artificial neural networks which recognize or even discover interesting patterns in the input data~ \cite{XX,haykin2009neural,nielsen2010nndl}. In a nutshell, the neural nets act as trainable functional mappings of multiple variables. To design even more powerful and intelligent machines requires one to quantify and extend the expressive power of the neural nets. However, there is a gap between the mathematical foundation and the real-world applications which are largely driven by the engineering practices~\cite{Goodfellow-et-al-2016-Book}, because it has long been a difficult endeavor to rigorously quantify the expressive power of neural nets over complicated datasets.

Insights into the physical rules governing the neural networks and typical datasets may offer an answer to the great success of deep learning and guide its more fruitful development in the future.
For example, statistical physics has a long-standing impact on machine
learning~\cite{PhysRevA.32.1007, haykin2009neural}, because both fields concern about collective behavior emerged from a large amount of microscopic degree of freedoms.
Moreover, as suggested by Mehta and Schwab~\cite{Mehta:2014ua}, there is a connection between the deep learning and the renormalization group.
Lin {\it et al.}~\cite{Lin2017}  also argued that the ``unreasonable success'' of deep learning can be traced back to the law of physics, which often
imposes symmetry, locality, compositionality, polynomial log-probability, and other properties on the input data. 


\begin{figure}[h!]
\begin{centering}
\includegraphics[width=1\columnwidth]{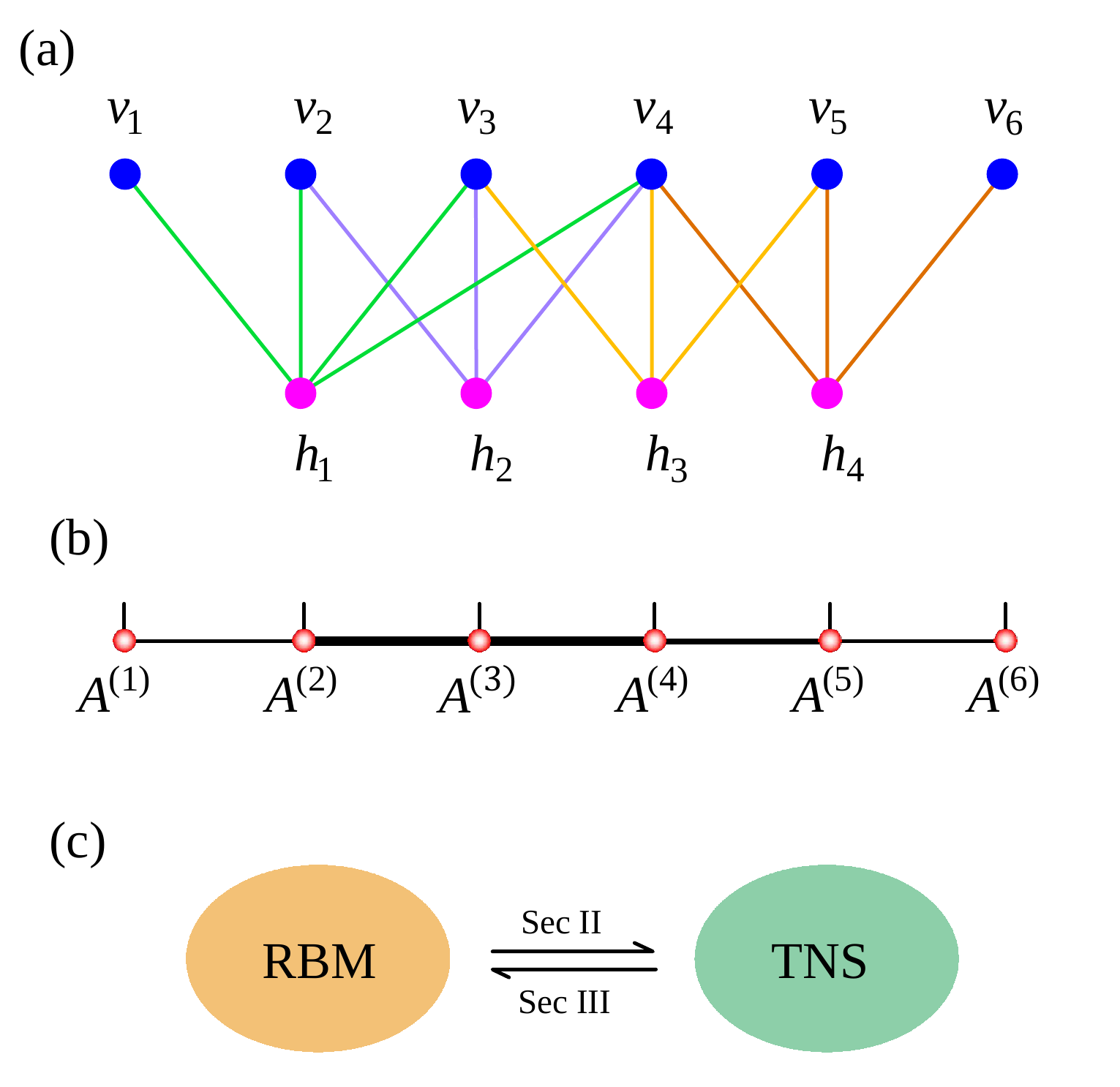}
\par\end{centering}
  \caption{(a) Graphical representation of the RBM defined by~\Eq{eq:RBM3_P}. The blue dots represent visible units $v$ and the magenta dots represent the hidden units $h$. They are coupled through the connections indicated by the solid lines. (b) MPS defined by~\Eq{eq:MPS}. Each red dot denotes a three-index tensor $A^{(i)}$. Throughout this paper, we use dots to represent units of RBM, and ball to represent tensors. Lines in the RBM denote connection weights while lines in the tensor network denote tensor indices. (c) RBM and TNS are two ways to parametrize multivariable functions. With unlimited resources d(number of hidden units or bond dimensions) both of them can represent any function to arbitrary accuracy. However, with limited resources they represent two independent sets with overlapping region. Detailed discussions on their relationship are given in Secs.~\ref{sec:RBM2TN} and \ref{sec:TN2RBM}.
\label{fig:setGraph}}
\end{figure}

Restricted Boltzmann machine (RBM) is a vivid example of the intrinsic connection between statistical physics and machine learning. An RBM is a special type of neural networks which can be better understood as an energy-based model. As illustrated in Fig.~\ref{fig:setGraph}(a), it consists of a set of interconnected visible and hidden binary variables. These variables are assumed to satisfy the Boltzmann distribution whose energy functional is defined by
\begin{equation}
 E\left(v, h\right) = -\sum_{i}a_{i}v_{i} - \sum_{j}b_{j}h_{j} - \sum_{i,j} v_{i}W_{ij}h_{j},
\end{equation}
where $v=\{v_{i}\}$ and $h=\{ h_{j} \}$ are the visible and hidden binary variables. We denote their number as $n_{v}$ and $n_{h}$
respectively. Parameters $a_{i}, b_{j}$ are the biases applied to the visible and hidden units, respectively.
$W_{ij}$ is the coupling matrix between these two units.

By integrating out the hidden units, the RBM represents  the marginal distribution of the visible variables (omitting an irrelevant normalization factor)
\begin{eqnarray}
    \Psi_\mathrm{RBM}\left(v\right) &= & \sum_{ h }e^{-E(v,h) } \nonumber  \\ & = & \prod_{i} e^{a_{i} v_{i}}\prod_{j}\left(1 + e^{b_{j}+ \sum_{i}v_{i}W_{ij}}\right). \label{eq:RBM3_P}
\end{eqnarray}
In the RBM there is no direct connection between the visible units. However the hidden units generate effective connections or interactions among them. By increasing the number of hidden units and connections, the RBM can in principle parametrize more complex functions of the visible units~\cite{Freund:1994tu,LeRoux:2008ex,Montufar}.

One can train an RBM by specifying its parameters such that the probability distribution of the visible units reproduces that of the input data~\cite{Hinton:2002ic,Tieleman:2008tj}.
The hidden units of a trained RBM may also reveal correlations in the data with appealing physical meanings. For example, in an RBM trained with a dataset of images containing handwritten digits, the connection weight contains the information about the pen stokes~\cite{RBMtutorial}. These learning features can be used either for discriminative tasks, such as pattern recognition, or for generative tasks, such as generating more samples according to the learned distribution. RBM has played an important role in the recent renaissance of deep learning~\cite{Hinton:2006kc, Hinton:2006dk}, because of its versatile abilities in feature extraction and dimensionality reduction of complex data sets.

%
%


Recently, RBM has attracted great attention in the field of quantum many-body physics. Carleo and Troyer~\cite{Carleo:2016vp} proposed an RBM inspired variational wavefunction to study quantum many-body systems at or away from equilibrium. Deng {\it et al.}~\cite{DLDeng2016} constructed exact RBM representations for several interesting topological states.  Torlai and Melko~\cite{Torlai:2016bm} trained an RBM to reproduce the thermodynamics of a statistical physics model. Huang and Wang~\cite{Huang:2016tf} used RBM as a recommender system to accelerate Monte Carlo simulation of quantum many-body systems. Liu {\it et al.}~\cite{PhysRevB.95.041101} reported similar ideas using classical spin models instead of the RBM.

These developments raise several critical questions about the expressive power of neural nets in the physics contexts. Is RBM more expressive than the standard variational wavefunctions of quantum states~\cite{Carleo:2016vp}? Can RBM efficiently describe the probability distribution of physical models at criticality~\cite{Torlai:2016bm,Huang:2016tf}?
Unfortunately, the existing universal approximation theorem~\cite{Freund:1994tu,LeRoux:2008ex,Montufar} and its further developments ~\cite{montufar2016hierarchical,NIPS2011_4380,Younes1996109}
 are not particularly instructive for practical purpose because they involve exponentially large resources, and it cannot be used as a guiding principle to solve practical physical or industrial problems.

In fact, the quest for more expressive wavefunction is central to quantum many-body physics. An ideal parametrization of wavefunction should accurately describe a quantum state with  exponentially large degree of freedoms with polynomial resources. Tensor network states (TNS)~\cite{Orus2014} are promising candidates to meet this demand.
Figure.~\ref{fig:setGraph}(b) shows one of the simplest TNS, the matrix product state (MPS)~\cite{MPSrepresentaion}, as an example. The MPS parametrizes a wavefunction of $n_{v}$ physical variables as,
\begin{equation}
\Psi_\mathrm{MPS}(v)  = \mathrm{Tr} \prod_{i} A^{(i)}[{v_{i}}] ,
\label{eq:MPS}
\end{equation}
where $A^{(i)}$ is a three-index tensor represented by a red dot in Fig.~\ref{fig:setGraph}(b). For a given value of $v_i$, which is represented by a dangling vertical bond,  $A^{(i)}[{v_{i}}]$ is a matrix. The dimension of this matrix is commonly called the virtual bond dimension of the MPS, indicated by the thickness of the horizontal bonds in Fig.~\ref{fig:setGraph}(b). Connecting these horizontal bonds is to take tensor contraction over all the virtual degree of freedoms.
By increasing the bond dimension, MPS can represent with increasing accuracy any complex multivariable functions~\cite{MPSrepresentaion}.

MPS representation is equivalent to the tensor train decompositions in the applied math community~\cite{oseledets2011tensor}. Similarly, one can connect higher order tensors to represent a physical state in a two dimensional network. This kind of TNS is named projected entangled pair state (PEPS)~\cite{peps}. A generalization of PEPS to include the entanglement of all particles in a larger unit cell is call projected entangled simplex state (PESS)~\cite{XiePRX2014}. In the past decades, solid theoretical understanding and efficient numerical techniques for TNS has been established. See~\cite{Verstraete:2008ko, Orus2014} for pedagogical reviews on TNS. Moreover, the application of TNS to classical systems also has a long history,  see \cite{Nishino1995, 2005PhRvL..95e7206M, TRGLevin} for example.

The physics community also has an answer to the ``unreasonable effectiveness'' of TNS. It relies on the entanglement area law~\cite{Eisert:2010hq}, which states that the entanglement entropy scales just linearly with the boundary size separating any two subsystems. The entanglement entropy~\cite{nielsen2010quantum} is a measure of the information content between these two subsystems. Many physical states of practical interests fulfill this area law~\cite{Eisert:2010hq}. It indicates that the degrees of freedom needed to describe a quantum state of physical interest is generally much less than the total degrees of freedom of the whole system. TNS are designed to efficiently represent these quantum states with relatively low entanglement entropy and have achieved remarkable successes in the past decades~\cite{2DDMRGreview}.

RBM and TNS share some similarities in their mathematical structures, especially expressed using the graphical language  in Figs.~\ref{fig:setGraph}(a,b). As for machine learning, Refs.~\cite{Lin2017, Goodfellow-et-al-2016-Book} also suggest that only a tiny fraction of the input data is of practical interests among infinite number of possible inputs. This motivates us to search for a guiding principle from the perspective of quantum information to quantify the expressive power of neural nets used for deep learning~\cite{LeCun:2015dt,Goodfellow-et-al-2016-Book} as well as for quantum and statistical physics problems~\cite{Carleo:2016vp, DLDeng2016, Torlai:2016bm, Huang:2016tf}.

In this paper, we present a general and constructive connection between RBM and TNS. With this correspondence, many concepts and techniques from deep learning and quantum physics can be exchanged.
By transforming an RBM to a TNS and exploiting its entanglement entropy bound, we can quantify the expressibility of RBM for quantum states, for statistical physics models, and for industrial datasets.
We also find the necessary and sufficient conditions for transforming a TNS into an RBM with a given structure, and show that RBM can serve as an efficient representation of quantum states.

This paper is arranged as follows.
In Sec.~\ref{sec:RBM2TN}, we present the algorithms to transform an RBM into an MPS or other kind of TNS and discuss their consequences.
In Sec.~\ref{sec:TN2RBM}, we present the sufficient and necessary conditions for a TNS to
have an RBM representation with a given architecture.
In Sec.~\ref{sec:ToricCode}, we illustrate the intimate connection between RBM and TNS by constructing the four-fold degenerate ground-state wave functions for the toric code model in both the RBM and TNS representations.
In Sec.~\ref{sec:application}, we discuss several applications of the established connection between RBM and TNS for physical and machine learning problems.
In Sec.~\ref{sec:discussion}, we discuss further implications of our results in a broader contexts of interdisciplinary research.
In Appendix~\ref{appendix:sufficient}, we present a sufficient condition to find the RBM representations for some specific TNS and illustrate with statistical Ising model and simple quantum states.
Finally, in Appendix~\ref{appendix:BM2TNS}, we discuss the equivalence between more general Boltzmann machines~\cite{hinton1986learning, DBM} and TNS.





\section{TNS representation of RBM \label{sec:RBM2TN}}

In this section we discuss the relationship between RBM and TNS via a constructive approach. 
An important application of the TNS representation for RBM is to provide an upper bound of the entanglement entropy it can capture. To estimate the bound one only needs structural information but not the detailed parametrization of the RBMs. We first present a simple and intuitive approach, then discuss more sophisticated approaches that can provide tighter or optimal bounds on the tensor bond dimensions. We provide code implementations of the mapping in~\cite{SM}.

Before proceeding, we first clarify a few notations about RBM.
First, in the standard machine learning applications and in Refs.~\cite{Torlai:2016bm,Huang:2016tf}, parameters $\{W_{ij},a_{i},b_{j}\}$ are assumed to be real and the corresponding RBM represents the probability distributions of input data.
However, in Refs.~\cite{Carleo:2016vp,DLDeng2016}, \Eq{eq:RBM3_P} is interpreted as the amplitude of quantum mechanical wave function, and $\{W_{ij},a_{i},b_{j}\}$ are generalized to the complex domain.
In this paper, we adopt the convention of Refs.~\cite{Carleo:2016vp,DLDeng2016}, and assume these parameters  to be complex.~\footnote{The references~\cite{Carleo:2016vp,DLDeng2016} used RBM units with $\pm 1$ instead of binary values. This amounts to a simple rescaling and offset of the weights and biases.}
Second, conventionally one views the RBM as an energy based model. While for our discussion about the expressibility of \Eq{eq:RBM3_P}, it is sufficient to view it as a function approximator such as a feed-forward neural net~\cite{marlin2010inductive, Huang:2016tf}. Third, although the standard RBM may have dense all-to-all connections between the visible and hidden units, for clarity we will illustrate the transformation using the RBM with sparse connections. Our result nevertheless holds generally and can be applied to RBMs with arbitrarily dense connections.

\begin{figure}
\begin{centering}
\includegraphics[width=1\columnwidth]{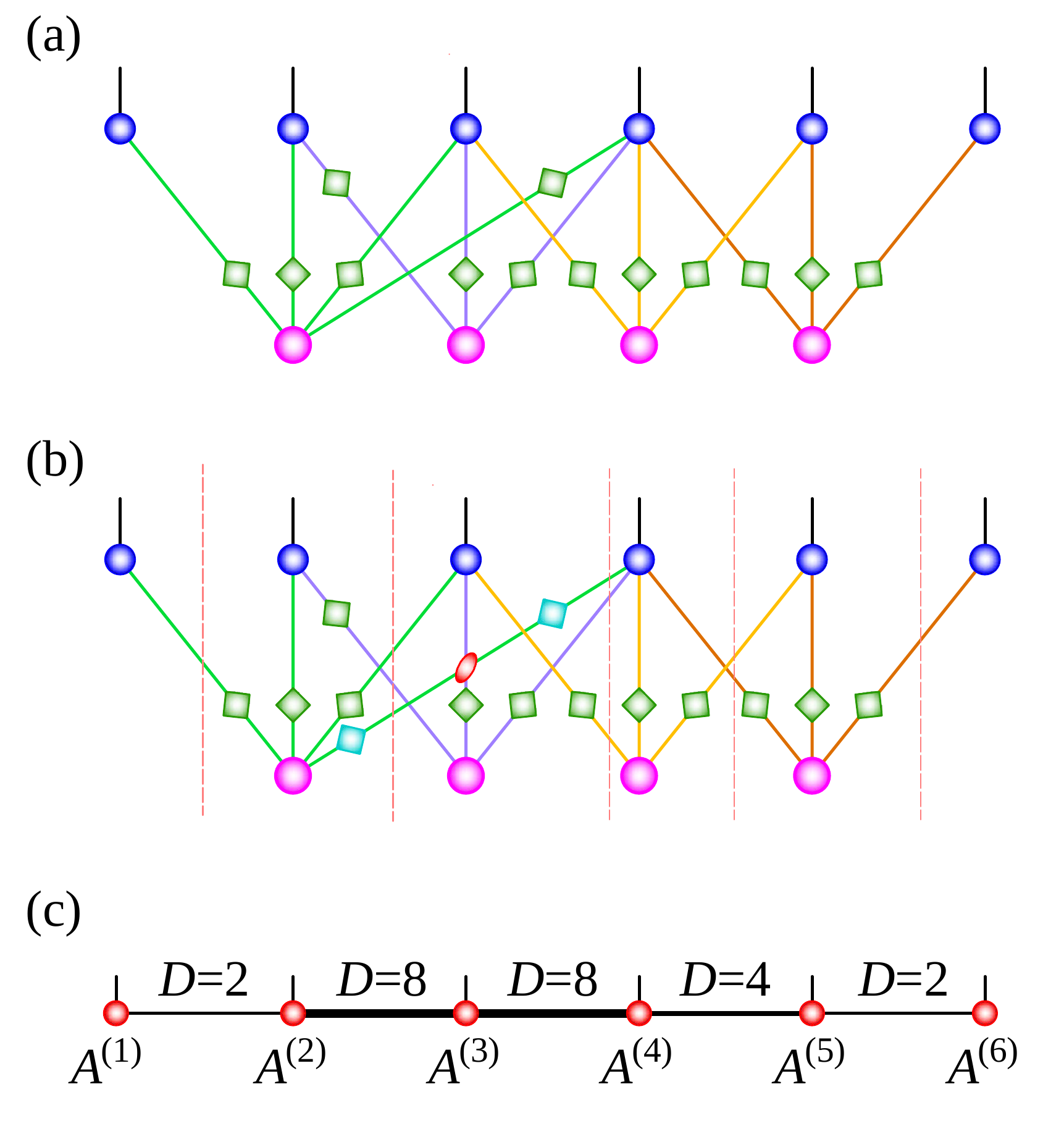}
\par\end{centering}
  \caption{Steps to map an RBM to an MPS. (a) A TNS representation of the RBM shown in Fig.~\ref{fig:setGraph}(a). The blue dots represent the diagonal tensors $\Lambda_{v}^{\left(i\right)}$ at the visible units and the magenta dots represent $\Lambda_{h}^{\left(j\right)}$ at the hidden units, defined by Eqs.~(\ref{eq:Lambdav}) and (\ref{eq:Lambdah}), respectively. The green squares represent matrix $M^{\left(ij\right)}$, defined in \Eq{eq:Mvh}. (b) The RBM is cut into $n_{v}$ (here $n_v=6$) pieces.  Associated with each long range connection, an identity tensor (red oval) is introduced to break $M^{\left(ij\right)}$ into two matrices, see Fig.~\ref{fig:swap}. (c) An MPS representation of the RBM obtained by contracting all hidden units and connection bonds in (b). The bond dimension of the MPS is determined by the number of bonds cut by the corresponding dashed vertical line. \label{fig:RBM2MPS}}
\end{figure}

\subsection{Direct Mapping of the RBM to MPS}
To start with, let us consider a simple way to map an RBM to an MPS. We summarize the procedure in Algorithm (\ref{alg:originalmethod}). As a concrete example, we consider the RBM defined in Fig.~\ref{fig:setGraph}(a). The first step is to convert this RBM into a TNS by representing the visible and hidden units as the physical and virtual variables while keeping the network structure unchanged. To do this, we decouple the Boltzmann weights into the terms defined on the vertices and bonds separately by introducing a diagonal tensor, $\Lambda_{v}^{\left(i\right)}$ or $\Lambda_{h}^{\left(j\right)}$, at each visible or hidden site, and a $2\times2$ matrix $M^{\left(ij\right)}$ on each bond linking $v_{i}$ and $h_{j}$:
\begin{eqnarray}
\Lambda_{v}^{\left(i\right)} & = & \mathrm{diag}\left(1, e^{a_{i}}\right), \label{eq:Lambdav} \\
\Lambda_{h}^{\left(j\right)} & = & \mathrm{diag}\left(1, e^{b_{j}}\right), \label{eq:Lambdah} \\
M^{\left(ij\right)} & = & \left(\begin{array}{cc}
1 & 1\\
1 & e^{W_{ij}}
\end{array}\right). \label{eq:Mvh}
\end{eqnarray}
This leads to the TNS representation of the RBM shown in Fig.~\ref{fig:RBM2MPS}(a).

\begin{figure}
\begin{centering}
\includegraphics[width=1\columnwidth]{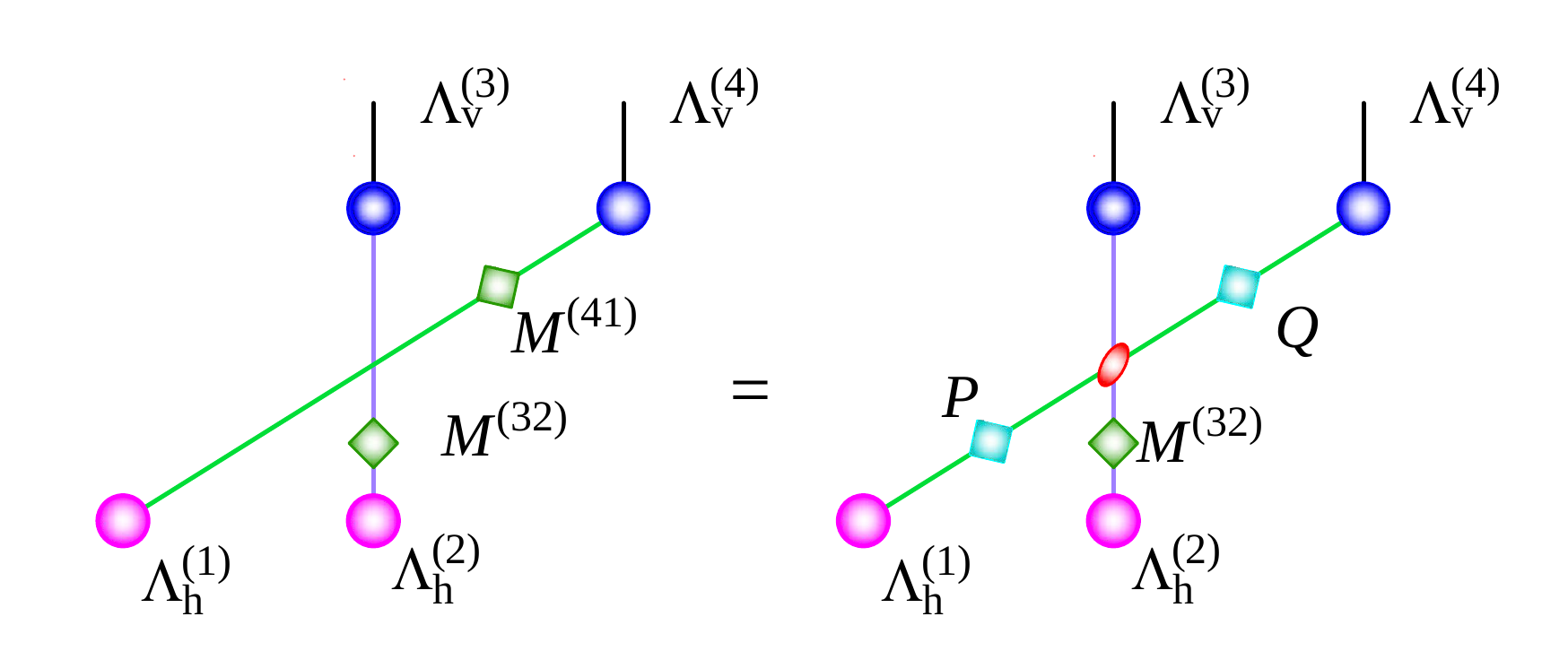}
\par\end{centering}
  \caption{Matrix $M^{\left(41\right)}$, represented by the green square, is decoupled into a product of two matrices, $P$ and $Q$, denoted by the two cyan squares. The red oval represents direct product of two identity matrix along two crossing directions.
 \label{fig:swap}
}
\end{figure}

The next step is to map the TNS in Fig.~\ref{fig:RBM2MPS}(a) to an MPS. We first cut the graphics into $n_v$ pieces, each containing a visible unit [Fig.~\ref{fig:RBM2MPS}(b)]. The assignment of the hidden units into these pieces can be arbitrary. Contracting all the internal variables within each piece is then equivalent to summing over the hidden units. The MPS, as shown in Fig.~\ref{fig:RBM2MPS}(c), is obtained by merging all the external connections between different pieces into the virtual bonds. The bond dimension of the MPS is indicated by the thickness of the virtual bonds, which is determined by  the number of connections merged.

Here we should pay more attention to the ``long-range'' connections that cross two or more vertical cuts. For the RBM shown in Fig.~\ref{fig:RBM2MPS}(a), the bond that connects $v_4$ and $h_1$ is the only long-range connection. In order to map it into the virtual bonds of MPS, we decouple the matrix defined on this connection, $M^{\left(41\right)}$, into a product of two arbitrary $2\times 2$ matrices, $P$ and $Q$, subject to the constrain $M^{\left(41\right)}=P\cdot Q$ (Fig.~\ref{fig:swap}). This separates effectively the long-range connection into two short ones, whose matrices are defined by $P$ and $Q$, respectively. These two matrices are then absorbed into the local tensors at $v_2$ and $v_3$, i.e., $A^{\left(2\right)}$ and $A^{\left(3\right)}$, respectively.
This long-range connection crosses the vertical bond at $v_3$, and consequently doubles the bond dimension of $A^{\left(3\right)}$.
In general, a long range connection will double the bond dimensions of all tensors it passes by. The dimension of MPS at a particular bond, $D$, is determined by the number of connections, $n$, one has to cut in order to bipartite the system at that bond, i.e. $D=2^{n}$.

It should also be noted that the MPS obtained from the above mapping process is not unique, because the geometrical structure of the hidden units with respect to the visible units can be arbitrarily arranged. No matter how the hidden units are arranged, these different MPS are equivalent. The local tensors obtained with the above approach generally also  contain redundant degrees of freedom. They can be gauged into a unique canonical form~\cite{canonicalClassical,canonicalQuantum} by taking canonical transformations for all local tensors. See Sec.~\ref{sec:redundancy} for more discussions on the redundancy of TNS and RBM representations.

\begin{algorithm}[H]
	\begin{algorithmic}[1]
		\caption{Direct mapping an RBM into an MPS.}
		\Require The connection weight matrix $W_{ij}$ and biases $a_{i}, b_{j}$. The matrix specifies the structure of the RBM. 
		\Ensure An MPS with each tensor $A^{(i)},i=1,2,\cdots,n_{v}$.  
		\State The RBM is cut into $n_v$ pieces $P_i$, $i=1,\cdots,n_v$. Each $P_i$ includes a visible unit $v_i$ and several hidden units.
		\For{$i=1,\cdots ,n_v$}
		\State Set $T_{i}= \varnothing$. \Comment Contains the tensors in $P_i$
		\State Construct $\Lambda_v^{(i)}$   according to Eq.~(\ref{eq:Lambdav}).
		\State Add $\Lambda_v^{(i)}$ to $T_i$.
		\For{ $h_j\in P_i$ } 
			\State Construct $\Lambda_h^{(j)}$ according to Eq.~(\ref{eq:Lambdah}).
			\State Add $\Lambda_h^{(j)}$ to $T_i$.
		\EndFor
		\EndFor
		\ForAll{$\{(v_i,h_j)|W_{ij}\neq0\}$}
		\Comment All the connections
			\State Construct $M^{(ij)}$ according to Eq.~(\ref{eq:Mvh}).
			\State Split $M^{(ij)}$ into products of matrices and add each matrix into the corresponding piece, see Fig.~\ref{fig:swap}.
		\EndFor
		\For{$i=1,\cdots, n_v$}
		\State Contract all the internal indices of the tensors within $T_i$, the result is $A^{(i)}$.
		\EndFor
		\label{alg:originalmethod}
	\end{algorithmic}
\end{algorithm}

\subsection{Optimal mapping of an RBM to an MPS}
\label{sec:method3}
The direct mapping method given in the last subsection although intuitive, is not optimal. Here we present a method to give the MPS representation with optimal bond dimension. 
RBM is an undirected probabilistic graphical model. For a graph model, we divide all variables into two sets $X$ and $Y$ which are conditionally independent if another set of variables $Z$ are given. This is written as 
\begin{equation}
X \perp Y | Z. 
\label{eq:perp}
\end{equation}
For the bipartition , we can identify the smallest set $Z$ such that Eq.~(\ref{eq:perp}) are satisfied. The degrees of freedom in $Z$ can be treated as virtual bond of an MPS after the translation. The size of $Z$, denoted by $|Z|$, determines the bond dimension between the variables $X$ and $Y$.
\begin{equation}
D=2^{\left|Z\right|}.
\end{equation}

Algorithm  (\ref{alg:optimalmethod}) lists the steps to translate an RBM to an MPS with optimal bond dimensions by employing such conditional independence property. We start from left and construct each tensor on the fly all along to the right side with the smallest virtual bond dimension. The virtual bonds of the resulting MPS represent the degrees of freedom of the visible or hidden units of the RBM.   

\begin{algorithm}[H]
\begin{algorithmic}[1]
	\caption{Transforming an RBM into an MPS with optimal bond dimensions,, see Fig.~\ref{fig:RBM2MPS_3}.}
		\Require The connection weight matrix $W_{ij}$ and biases $a_{i}, b_{j}$. The matrix specifies the structure of the RBM. 
		\Ensure An MPS with each tensor $A^{(i)},i=1,2,\cdots,n_{v}$.  
	
	\State $\mathscr{G}_{s}=\{(i, j)|W_{ij}\neq0\}$
	\Comment Graph formed by connected units
	\State $\mathscr{H}_{s}=\{j | (i , j)\in \mathscr{G}_{s} \}$
	\Comment All hidden units
	\State $Z^{\prime} = \varnothing $ 
	\Comment The degrees of freedom of the left virtual bond
		\For{$i=1,\cdots, n_v$}
		\State $\mathscr{G}_{t}=\varnothing$
		\Comment Connections to be counted in tensor $A^{(i)}$
				\State $\mathscr{H}_{t}=\varnothing$
				\Comment Hidden units to  be traced in tensor $A^{(i)}$
		\State $X=Z^{\prime} \cup \{v_i\}$
		\State $Y=\{v_{i+1},v_{i+2},\cdots , v_{n_v}\}$
		\Comment The remaining physical degrees of freedom 
		\State Find a minimal set $Z$ such that  $X\perp Y | Z$ on the graph $\mathscr{G}_{s}$. 
		\For { $j  \in \mathscr{H}_{s}$}
			\If{$h_{j}$ is not connected to $(Y\setminus Z)$}
				\State Move $j$ from $\mathscr{H}_{s}$ to $\mathscr{H}_{t}$. 
				\Comment Variable $h_{j}$ will be traced out in tensor $A^{(i)}$.
			\EndIf
		\EndFor
		\For {$(k, j)\in \mathscr{G}_{s}$}
			\If {$v_{k}$ and $h_{j}$ belongs to $X \cup Z \cup \mathscr{H}_{t}$}
				\State Move $(k,j)$ from $\mathscr{G}_{s}$ to $\mathscr{G}_{t}$.
				\Comment The $(v_{k},h_{j})$ interaction will be  included in tensor $A^{(i)}$.
			\EndIf
		\EndFor
		\State $A^{(i)}_{Z^{\prime},Z}[v_{i}]=\sum_{ \{h_{j\in \mathscr{H}_{t} }\}}e^{ a_{i} v_{i}+\sum_{(k,j)\in \mathscr{G}_{t}}v_{k}W_{kj}h_j+\sum_{j\in \mathscr{H}_{t}}b_j h_j}$
		\State $Z^{\prime} \gets Z$
	\EndFor
	\label{alg:optimalmethod}
\end{algorithmic}
\end{algorithm}

\comments{
1.Suppose we are constructing tensor $A^{(i)}$. The left side are $\{v_1,\cdots,v_i\}$ and right side are $\{v_{i+1},\cdots v_{n_v}\}$.

2. Keep in mind the interactions which have not yet been counted, noted as $I=\{(v,h)\}$. For RBM, the interaction is given by visible and hidden units pair. 

3. Set $X$ includes last virtual bond variables and current physical variable. Set $Y$ includes all the visible variables in the right.

4. Search for the minimum size set $C$, so that it satisfies Eq.~\ref{eq:perp}. $C$ determines the virtual bonds.

5. For each hidden units in $\{v,h\}$ pairs of $I$, if it have no interactions with variables in $B-C$. It is traced over as inner freedom in the tensor.

6. Go back to 1
}

\begin{figure}
	\begin{centering}
		\includegraphics[width=1\columnwidth]{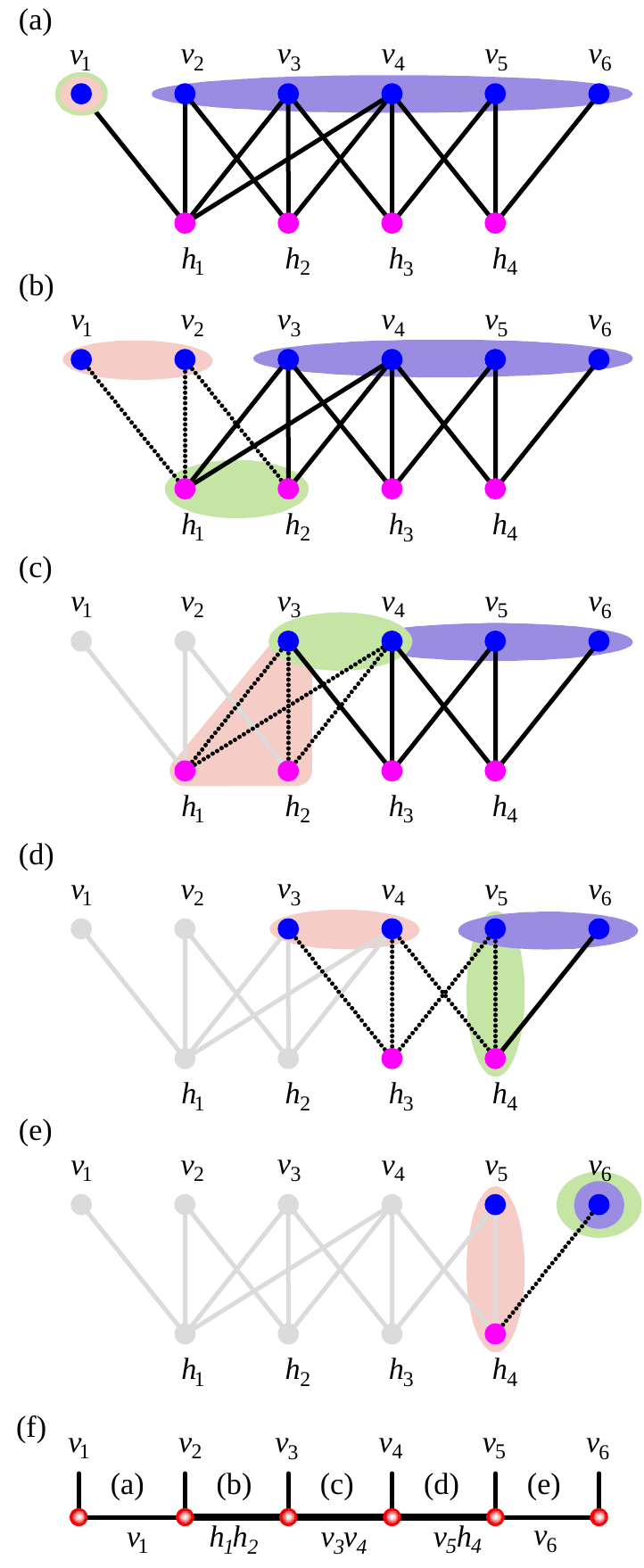}
		\par
	\end{centering}
	\caption{The optimal mapping from the RBM to the MPS. (a)-(e) shows each step of the construction. The set $X$ is denoted by pink ellipse and $Y$ in purple ellipse. $Z$ is the minimal set satisfying $X \perp Y|Z$, denoted in light green ellipse. When the variables in $Z$ are given, the RBM function factorize into product of functions of variables in $X$ and $Y$. The degrees of freedom in $Z$ is represented by the virtual bond of the MPS. The connection lines in gray represent interactions counted by previous tensors. The doted line indicates those under consideration at the current step, denoted by $\mathscr{G}_t$  in Algorithm~(\ref{alg:optimalmethod}). The constructed tensor in (a)-(e) are given in (f). }
		\label{fig:RBM2MPS_3}
\end{figure}

To illustrate the mapping algorithm, we take the RBM in Fig.~\ref{fig:setGraph}(a) as an example. 
We start in Fig.~\ref{fig:RBM2MPS_3}(a) by considering $X = \{v_1\}$ and $Y =\{v_2,v_3, \cdots, v_6\}$. It is straightforward to see that $Z$ can either be $\{v_1\}$ or $\{h_1\}$ to satisfy the conditional independence $X \perp Y | Z$. Suppose we take $Z=\{v_1\}$, $\mathscr{G}_t = \mathscr{H}_t =\varnothing$, the first tensor $A^{(1)}$ can therefore be chosen as an identity tensor which copies the visible variables $\{v^{(1)}\}$ to the right virtual bond of the matrix $A^{(1)}$. 

In Fig.~\ref{fig:RBM2MPS_3}(b), we select $X=\{v_1, v_2\}$ and $Y=\{v_3,\cdots, v_6\}$. 
We can find that the smallest $Z$ can be any of $\{v_1,v_2\}$,$\{v_2,h_1\}$,$\{v_3,v_4\}$ or $\{h_1,h_2\}$. We choose $Z=\{h_1,h_2\}$ for example. Note that the set $Z$ can contain both visible and hidden units. The $\mathscr{G}_t$ in Algorithm~(\ref{alg:optimalmethod}), denoted by the dashed lines in Fig.~\ref{fig:RBM2MPS_3}, contains all the connections considered by the tensor $A^{(2)}$. $\mathscr{H}_t=\varnothing$. The left bond of $A^{(2)}$ is the same as the right bond of $A^{(1)}$. And the right bond of $A^{(2)}$ consists of $Z=\{h_1,h_2\}$. 

And we go on to Fig.~\ref{fig:RBM2MPS_3}(c), where we denote all the interactions which have been taken account as gray lines. In this step, the set $X=\{h_1,h_2\} \cup \{v_3\}$, line $7$ of Algorithm  (\ref{alg:optimalmethod}). There are several choices of $Z$ to reach the minimum sizse $\left|Z\right|=2$. Here, we choose $Z=\{v_3,v_4\}$. Therefore, $\mathscr{G}_t$ consists of all the connections between $\{h_1,h_2\}$ and $\{v_3,v_4\}$. The corresponding interactions are considered by $A^{(3)}$, whose  right bond consists of $Z=\{v_3,v_4\}$. $\mathscr{H}_t=\varnothing$, no hidden units needs to be traced out.

Finally, as we come to Fig.~\ref{fig:RBM2MPS_3}(d), $Z$ includes $v_5$ and $h_4$. We find $\mathscr{G}_t=\{h_3\}$ has no interaction with the variables in the set $Y$ when $Z$ is given. So $h_3$ is traced out when constructing $A^{(4)}$. 

Each connection line of the original RBM is considered only once during the construction of the MPS. In this way, we can obtain all the six tensors. We label the bond degrees of freedom in Fig.~\ref{fig:RBM2MPS_3}(f).  
Note that even if one is not running the algorithm numerically, one can still obtain the optimal bond dimension of the corresponding MPS. Moreover, this method can be used in for general undirected graphical model. 

\subsection{Implication of the RBM to MPS mapping}

The connection between TNS and RBM suggests that we can use the knowledge of TNS, especially entanglement entropy, to quantify the expressibility of RBM.
Let us divide the visible units into two parts, denoted as $X$ and $Y$, respectively.
The entanglement entropy of a function $\Psi$ (which can either be an RBM or an MPS) between these two subsystems is then given by~\cite{nielsen2010quantum}
\begin{equation}
S = -\mathrm{Tr}(\rho \ln \rho),
\label{eq:Sent}
\end{equation}
where $\rho$ is the reduced density matrix defined by
\begin{equation}
\rho = \sum_{v_{Y}} \Psi^{\ast}\left( v_X^{\prime},v_Y\right)\,\Psi\left(v_X,v_Y\right).
\label{eq:rho}
\end{equation}
The matrix is spanned by the visible degrees of freedom in $X$, while $v_Y$ contains all the visible units in $Y$. The entanglement entropy characterizes the information content of $\Psi$, and can be viewed as a proxy of correlations between $X$ and $Y$. In case $X$ and $Y$ are completely disentangled, the entanglement entropy \Eq{eq:Sent} is zero. While if there are only short range correlations, the entanglement entropy should depend only on the size of the interface separating $X$ and $Y$, which is small in comparison with the full volume of the system~\cite{PhysRevLett.100.070502}. The entanglement entropy of MPS can be readily calculated. The maximal entanglement entropy is bounded by the logarithm of the bond dimension, i.e. $\ln D$~\cite{2011AnPhyS}.

\begin{figure}
\begin{centering}
\includegraphics[width=1\columnwidth]{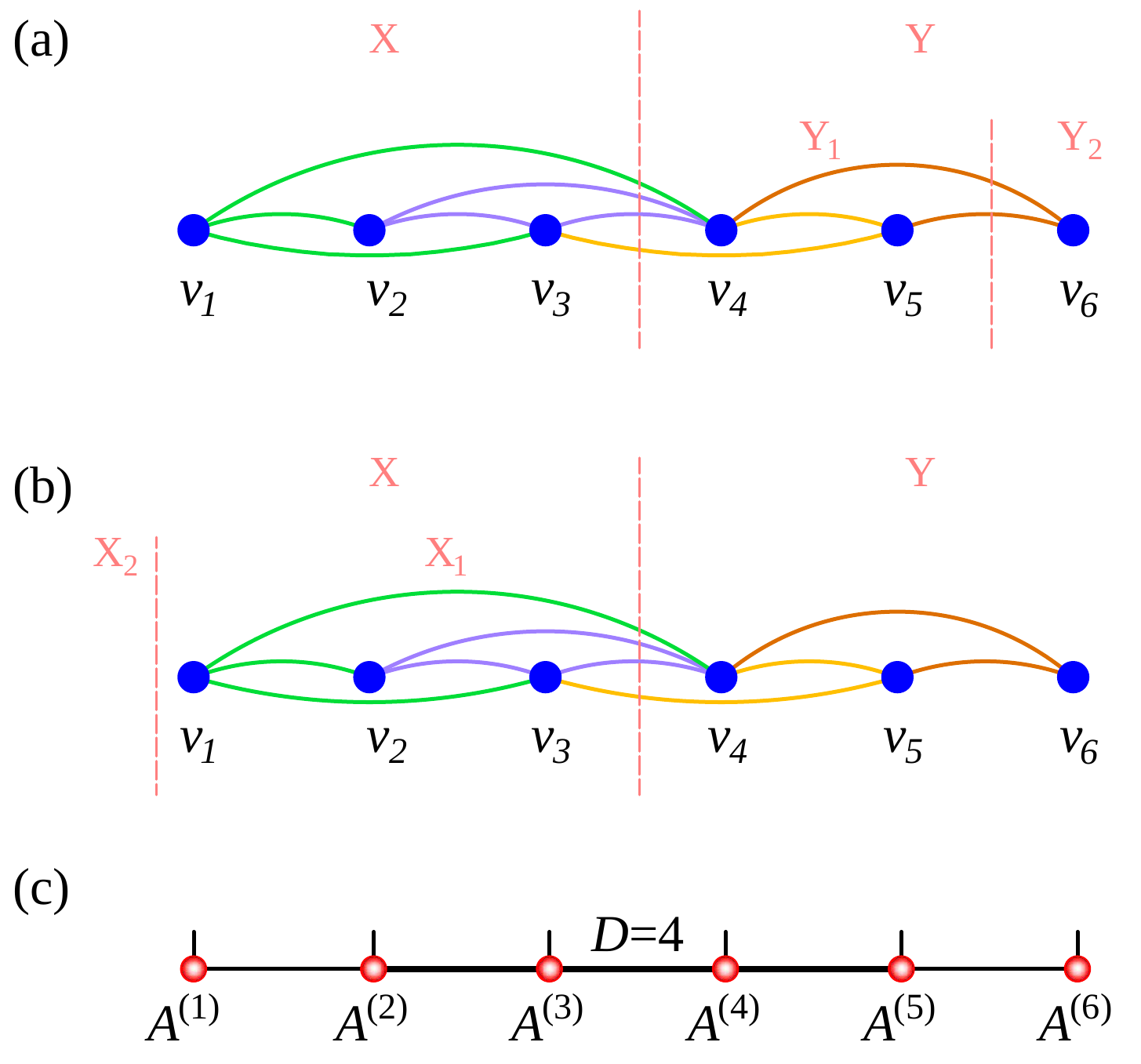}
\par\end{centering}
  \caption{(a) RBM with all hidden units traced out. The Arcs represent the connections between the visible units mediated by the hidden units. The system is divided into two subsystems, $X$ and $Y$. $Y$ is then further divided into two parts, $Y=Y_{1}\cup Y_{2}$, where $Y_{1}$ contains all the units that directly connect to the units in $X$. (b) Alternatively, $X$ is divided into two parts $X=X_{2}\cup X_{1}$, where $X_{1}$ contains units that connect directly to the units in $Y$. (c) The resulting MPS with smaller bond dimensions in comparison with that shown in Fig.~\ref{fig:RBM2MPS}(c). \label{fig:newMethod}}
\end{figure}

To better assess the expressive power of an RBM one needs  to find an equivalent MPS representation with the smallest possible bond dimensions. However, the simple and intuitive approach illustrated in Fig.~\ref{fig:RBM2MPS}  just provide an upper bound of the bond dimension which is generally higher than what is needed. For example, the bond dimension of the second bond, $D=8$, in Fig.~\ref{fig:RBM2MPS}(c) is more than enough to capture the entanglement since there are only two visible units on its left, whose total degrees of freedom just equals four. Below we present more sophisticated approaches to obtain a tighter bound on the bond dimensions. This improved approach is independent on the assignment of the hidden units.

Figure~\ref{fig:newMethod}(a) shows an RBM after tracing out all the hidden units.
The arcs indicate the interactions between the visible units mediated by the hidden units. If we separate the visible units into two parts, $X$ and $Y(=Y_{1}\cup Y_{2})$, where the interface region $Y_{1}$ contains all the visible units that directly link to $X$, and $Y_{2}$ the rest units, the RBM function, \Eq{eq:RBM3_P}, can then be expressed as
\begin{equation}
\Psi_\mathrm{RBM}\left(v\right) =\psi\left(v_X, v_{Y_1}\right)\phi\left(v_{Y_1}, v_{Y_2}\right). \label{eq:WaveABC}
\end{equation}
Once the visible units in $Y_{1}$ are fixed, this RBM  becomes a direct product of the visible units in $X$ and $Y_2$.
The rank of this function~\cite{ranktensor, hitchcock-sum-1927}, or the entanglement entropy between $X$ and $Y$, is therefore determined by the total number of visible units in the interface region $Y_{1}$, denoted by $|Y_{1}|$. Hence, the bond dimension of the MPS on the bond separating $X$ and $Y$ is simply given by $D = 2^{|Y_{1}|}$.

Alternatively, one can also divide $X$ into two parts $X=X_{2}\cup X_{1}$ by including all the units that have direct connections with $Y$ in $X_1$ and the rest units in $X_2$ [Fig.~\ref{fig:newMethod}(b)]. Following the argument given above, it can be shown that the entanglement entropy between $X$ and $Y$ is also bounded by the number of units in $X_1$. Thus, the entanglement entropy between $X$ and $Y$ is bounded by $S_\mathrm{max}= {\min (|X_1|, |Y_1|)}\ln2 $. \footnote{Similarly, when modeling the probability using probabilistic graphical models, the upper bound of classical mutual information is given by the size of the interface~\cite{PhysRevLett.100.070502}.}

Figure~\ref{fig:newMethod}(c) shows the MPS obtained with this approach. The bond dimensions from left to right are $D=2,4,4,4,2$, respectively. They are tighter bounds on the bond dimensions of MPS compared to Fig.~\ref{fig:RBM2MPS}(c). More generally, an even tighter bound on bond dimension can be obtained by fixing a minimal number of units, no matter they are visible or hidden, such that the RBM separates into a bipartite product state. To directly construct the MPS with the optimal smallest bond dimensions, we represent the degrees of freedom in the interface region using the virtual bonds of the MPS. Programming codes to implement these RBM to MPS mapping algorithms are given in~\cite{SM}.


The bond dimensions of the resulting MPS control the maximal entanglement entropies between the visible units. Therefore, the entanglement entropy provides a rigorous quantification on the expressive power of an RBM  solely based on its architecture. Estimating these bounds can be done efficiently with the provided codes~\cite{SM}. Moreover, canonizing the MPS may further reduce the bond dimensions by removing all unnecessary degrees of freedom and give precise value of the entanglement entropies.


Similarly, one can map an RBM into a PEPS~\cite{peps} by arranging the visible units on a two-dimensional array. A similar procedure was used in \cite{PhysRevLett.105.010502} to map a multi-scale entanglement renormalization ansatz~\cite{MERA} into a PEPS. This is particularly useful if the original dataset represented by the RBM, for example the pixels of image, is defined on a two-dimensional grid.

In general, if the number of units along any direction of the interface region is bounded by $m$, then the upper bound of the entanglement entropy should scale as
\begin{equation}
S_{\mathrm{max}} \sim m V^{(d-1)/d}, \label{eq:S_area}
\end{equation}
where $d$ is the spatial dimension on which the TNS is defined, and $V$ is the volume of the system. Thus, the maximum entanglement of a sparsely connected RBM~\cite{DLDeng2016} satisfies the area law. However, for a densely connected RBM, the interface region extends to the whole system and $m \sim V^{1/d}$, therefore $S_{\mathrm{max}}\sim V$ scales linearly with the subsystem volume.
This suggests that the dense RBM can provide a compact representation for a highly entangled quantum state that does not satisfy the entanglement area law. The number of parameters in this dense RBM just scales polynomially with the system size. However, to
describe this state using an MPS or a PEPS, the number of parameters needed scales exponentially with the system size~\cite{MERA, branchingMERA}. This provides an entanglement entropy justification for the variational calculation of quantum systems using RBM functions~\cite{Carleo:2016vp}.  Section~\ref{sec:shiftinvariantRBM} presents a detailed analysis of the tensor representation of the state used in Ref.~\cite{Carleo:2016vp}.

The mapping from RBM to TNS is valid more generally and can be extended to Boltzmann machines without the bipartite graph restriction. In Appendix~\ref{appendix:BM2TNS}, we discuss the general equivalence between Boltzmann machines~\cite{hinton1986learning} and TNS.

\section{RBM representation of TNS: sufficient and necessary conditions \label{sec:TN2RBM} }

We now address the reverse question about how to transform a TNS into an RBM with a given architecture.
Here only a given architecture of RBM is considered because otherwise one can reproduce any function using an RBM with exponentially large resources~\cite{Freund:1994tu,LeRoux:2008ex,Montufar}. Again we present a constructive approach to determine the RBM parameters for a TNS.
In Appendix~\ref{appendix:sufficient}, a sufficient condition for mapping a TNS to an RBM is discussed and demonstrated with the statistical Ising model and the cluster state~\cite{ClusterState}.

\begin{figure}
\begin{centering}
\includegraphics[width=1\columnwidth]{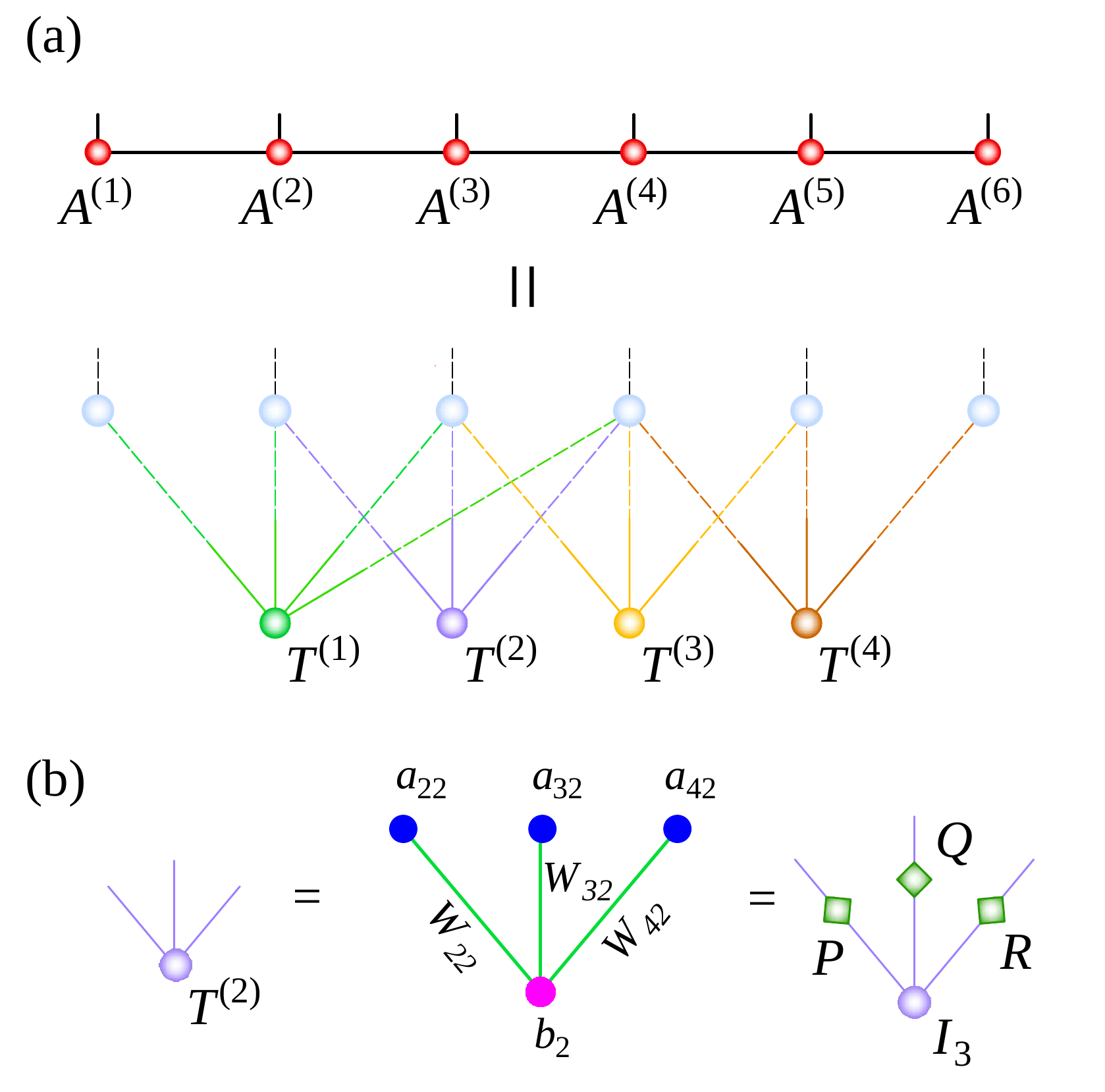}
\par\end{centering}
\caption{Graphical representation of (a) \Eq{eq:logTr} and (b) Eqs.~(\ref{eq:T3Eq},\ref{eq:TeqABC}).
\label{fig:MPS2RBM}}
\end{figure}

Let us take the 6-site MPS shown in Fig.~\ref{fig:setGraph}(b) as an example to show how to parameterize it into an RBM with the architecture shown in Fig.~\ref{fig:MPS2RBM}(a).
The hidden layer contains $n_{h}=4$ units, which factorize the MPS into a product of 4 tensors, one $2\times2\times2\times2$ tensor, $T^{(1)}$, defined at $h_1$, and three $2\times2\times2$ tensors, $T^{\left(2,3,4\right)}$, defined at the remaining three hidden units.
Requiring this product to equal the MPS, we have
\begin{equation}
\mathrm{Tr} \prod_{i} A^{\left(i\right)}\left[v_{i}\right]
  =  T_{v_{1}v_{2}v_{3}v_{4}}^{\left(1\right)}T_{v_{2}v_{3}v_{4}}^{\left(2\right)}T_{v_{3}v_{4}v_{5}}^{\left(3\right)}T_{v_{4}v_{5}v_{6}}^{\left(4\right)}. \label{eq:logTr}
\end{equation}
Taking the logarithm of this equation, we obtain $2^{n_v}=64$ linear equations for the $2^{4}+2^{3}+2^{3}+2^{3}=40$ tensor elements.
These equations are over determined since the number of parameters is less than the number of equations.
In order to map a TNS to an MPS, these equations must have an unique solution.
If these equations have no solution, then one has to change the architecture of RBM. One can, for example, increase the number of parameters by increasing the number of hidden units and/or the number of connections of the RBM.
This set of linear equations may become underdetermined if the number of parameters in the RBM becomes larger than the number of equations.

If \Eq{eq:logTr} has an unique solution, we then need to decompose each tensor into an RBM with just one hidden unit. For example, for the tensor shown in Fig.~\ref{fig:MPS2RBM}(b), i.e. $T^{(2)}$, we should decouple it as
\begin{equation}
T_{v_{2}v_{3}v_{4}}^{\left(2\right)}=\sum_{h_{2}\in\{0,1\}}e^{  h_{2}b_{2} + \sum_{i\in\{2,3,4\}} v_{i} \left(W_{i2}h_{2}+a_{i2}\right)},  \label{eq:T3Eq}
\end{equation}
where $a_{ij}$ is a partial bias of the $i$'th visible unit contributed by the $j$'th hidden unit.
The bias of the $i$'th visible unit is given by the sum of all partial biases contributed by the hidden units connecting this visible unit, $a_{i}=\sum_{j}a_{ij}$.
For this 3-index tensor, 7 parameters need to be determined from $2^{3}=8$ equations [Fig.~\ref{fig:MPS2RBM}(b)].
The number of parameters grows linearly with the order of $T$, but the number of equations grows exponentially instead.
In general, \Eq{eq:T3Eq} is over-determined, and it has solution only in special cases.

In practice, the \Eq{eq:T3Eq} can be considered as the tensor rank decomposition (CP decomposition)~\cite{ranktensor, hitchcock-sum-1927} of $T^{(2)}$,
\begin{equation}
T_{v_{2}v_{3}v_{4}}^{\left(2\right)}= \sum_{h_{2}\in\{0,1\}}P_{v_{2}h_{2}}Q_{v_{3}h_{2}}R_{v_{4}h_{2}}.\label{eq:TeqABC}
\end{equation}
The rank of $T^{(2)}$ is 2 because the hidden unit ($h_2$ here) is a binary variable.
$P, Q, R$ are all $2\times2$ matrices because the visible units are all binary numbers.~\footnote{If the rank of tensor is higher than two, then one has to enlarge the basis dimension of hidden units from 2 to a larger number. }
For a $2\times2\times2$ tensor, a rank-2 decomposition always exists in complex field and a rank-4 decomposition in real field~\cite{landsberg2012tensors}. 
However, for an arbitrary tensor, it is difficult to determine its rank~\cite{ranktensor}.
The high order singular value decomposition~\cite{kolda2009tensor} gives the lower bound of the tensor rank as the dimension of the core tensor. If it is larger than $2$, the binary condition of the hidden unit cannot be satisfied.

After the decomposition \Eq{eq:TeqABC}, we can further decouple each matrix into a product of three matrices according to Eqs.~(\ref{eq:Lambdav}-\ref{eq:Mvh}). For example, we can express matrix $P$ as
\begin{equation}
P= \left(\begin{array}{cc}
p & q\\
r & s
\end{array}\right) = p\left(\begin{array}{cc}
1\\
 & \frac{r}{p}
\end{array}\right)\left(\begin{array}{cc}
1 & 1\\
1 & \frac{ps}{qr}
\end{array}\right)\left(\begin{array}{cc}
1\\
 & \frac{q}{p}
\end{array}\right).\label{eq:AeqLWL}
\end{equation}
Comparing to Eqs.~(\ref{eq:Lambdav}-\ref{eq:Mvh}), we obtain
\begin{eqnarray}
W_{22} & = & \ln\frac{ps}{qr},\label{eq:W}\\
a_{22} & = & \ln\frac{r}{p},\label{eq:La}\\
b_{22} & = & \ln\frac{q}{p}.\label{eq:Lb}
\end{eqnarray}
Similar to $a_{ij}$, $b_{ij}$ is a partial bias to the $j$'th hidden unit imposed by the $i$'th visible unit. The bias of the $j$'th hidden unit is given by $b_{j} =\sum_{i} b_{ij}$.
In this way, each tensor of \Eq{eq:logTr} is written in the RBM form, for example.

Thus the necessary and sufficient condition for mapping an MPS to an RBM is that both
Eq.~(\ref{eq:logTr}) and Eq.~(\ref{eq:T3Eq}) have unique solutions.
In the case of $n_{h}=1$, \Eq{eq:logTr} merely rephrases the MPS as the wavefunction itself. The rank-2 decomposition of the tensor, similar to \Eq{eq:T3Eq}, is generally more difficult to satisfy. By increasing the number of hidden units and connections, one can increase the number of parameters of RBM to ensure both Eq.~(\ref{eq:logTr}) and Eq.~(\ref{eq:T3Eq}) to have solutions. This agrees with the mathematical results stating that RBM can represent any function by employing an exponentially large number hidden units and connections~\cite{Freund:1994tu,LeRoux:2008ex,Montufar}.

In practice, a convenient way to quickly check whether a state has a particular RBM representation is to consider the factorization property defined by \Eq{eq:WaveABC}, namely to examine whether a TNS can be factorized into a product state by fixing a sequential of visible units.

Using this simple approach, we can actually show that the spin-1 Affleck-Kennedy-Lieb-Tasaki (AKLT) state~\cite{AKLTstate} cannot be represented as an RBM (with ternary visible units) with only short-range connections because of the existence of hidden string order. Expressed in the $(S^z)^{\otimes n_{v}}$ basis, each component of the wave function looks like ``$\mathsf{ +\ 0\,\ 0\, -\, 0\, + -\, 0\,\ 0\,\ 0\, + }$'', where $\left( +\, ,\,0\, ,\,-\right)$ represent the eigenstates with $S^z=\left(1,0,-1  \right)$. There is a hidden antiferromagnetic order with arbitrary number of ``$\mathsf{0}$''s inserted. Even if we fix a sequence of visible units  in the middle to be ``$\mathsf{0}$'', the state is still combination of ``$\mathsf{+\ 0\,\ 0\,\ 0\cdots 0\,-}$'' and ``$\mathsf{-\,0\,\ 0\,\ 0\cdots 0\, +}$'', which cannot be simply expressed as a product state, or an RBM with just local connections.
On the other hand, the AKLT state can be written as a $D=2$ MPS \cite{AKLTstate, Orus2014}. This example shows that the entanglement entropy is not the only variable that quantifies the expressive power of RBM. 

\section{Example: RBM representation of the toric code ground states \label{sec:ToricCode}}

As a concrete example of the TNS-RBM transformation introduced in the preceding section, we construct the RBM representation of the toric code ground states from the corresponding PEPS wavefunctions~\cite{KitaevToric}. More examples, including the Ising model and the cluster state~\cite{ClusterState}, are given in the Appendix~\ref{appendix:sufficient}.

The toric code~\cite{KitaevToric} is one of the simplest models whose ground states are topological ordered~\cite{QFTWen} and holds the promise for quantum computation~\cite{KitaevSurface}. It is defined by the Hamiltonian
\begin{eqnarray}
\mathcal{H} & =& -\sum_{+}\mathcal{A}_{+}-\sum_{\square}\mathcal{B}_{\square},\label{eq:TCH} \\
\mathcal{A}_{+} & = & \prod_{i\in +}\sigma_{i}^{z}, \quad
\mathcal{B}_{\square}  = \prod_{i\in \square}\sigma_{i}^{x},\label{eq:AB}
\end{eqnarray}
where $\sigma^{x}_{i}$ and $\sigma^{z}_{i}$  are Pauli matrices defined on the horizontal and vertical links of the square lattice. $\mathcal{A}_{+}$ consists of the product of four $\sigma^{z}_{i}$ operators connecting to each vertex denoted by $+$, and $\mathcal{B}_{\square}$ consists of the product of four $\sigma^{x}_{i}$
operators on each minimal square plaquette $\square$. All the $\mathcal{A}_{+}$ and $\mathcal{B}_{\square}$ operators commute with each other.

The toric code on a torus has four topologically degenerate ground states.
Deng {\it et al.}~\cite{DLDeng2016} has already found an RBM representation for one
of them. We now present the RBM representations for all the four ground states using the approach introduced in the preceding section based on their PEPS representations~\cite{ToricPeps,Orus2014}.

\begin{figure}
\begin{centering}
\includegraphics[width=1\columnwidth]{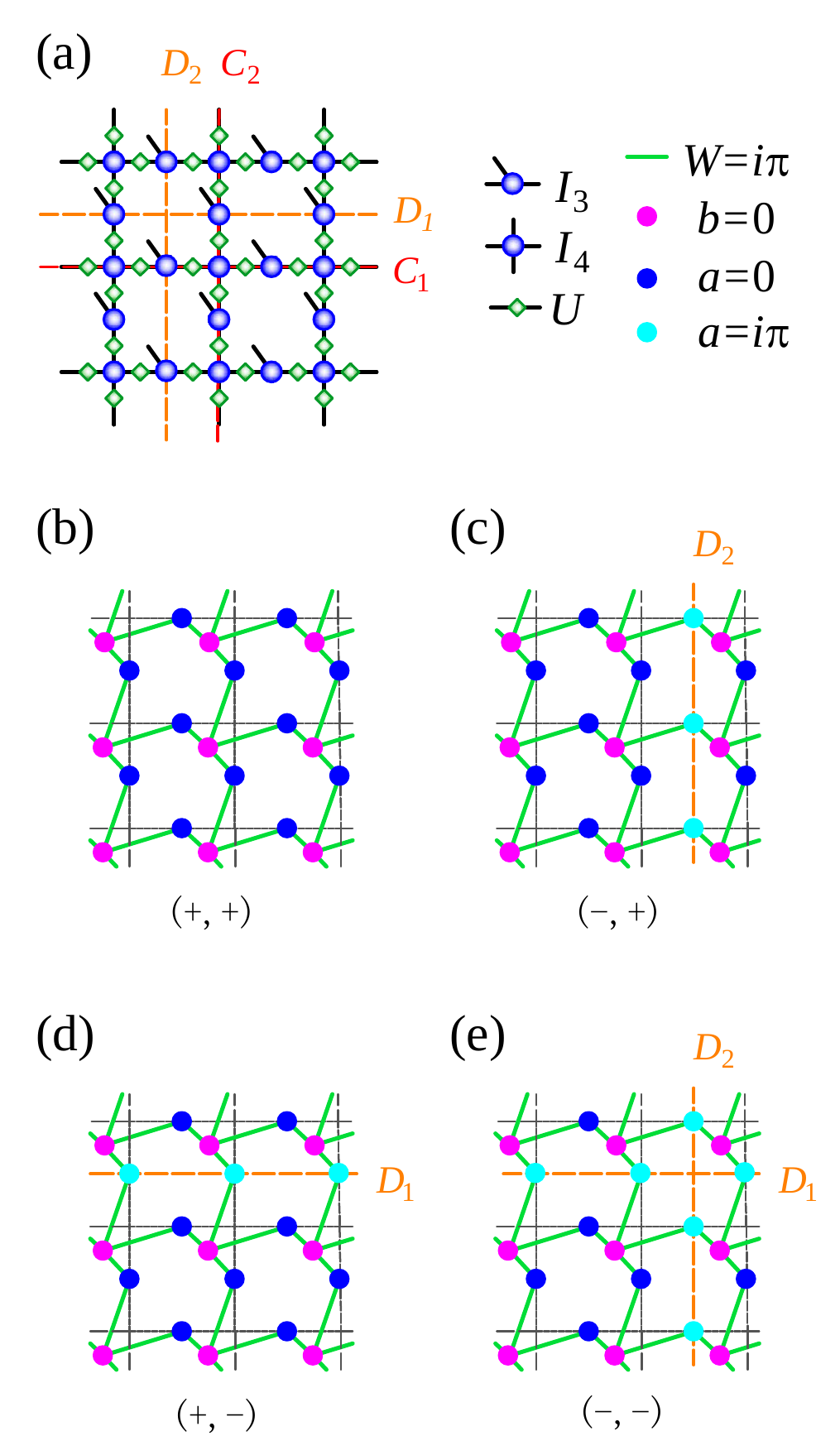}
\par\end{centering}
  \caption{(a) TNS and (b) its corresponding RBM representations of the toric code ground state at the $(+,+)$-topological sector on the square lattice. The dashed red and orange lines are the paths of Wilson loops used in \Eq{eq:X12} and \Eq{eq:Z12}, respectively.
  (c-e) RBM representations of the toric code ground states in the other three topological sectors. The cyan dots denote the visible units with bias $a=\mathrm{i}\pi$ . \label{fig:Toric}}
\end{figure}

We start from the PEPS representation for one of the ground states of the toric code model
~\cite{ToricPeps,Orus2014} shown in Fig.~\ref{fig:Toric}.
The PEPS consists of three kinds of local tensors of dimension 2: a 4-index identity tensor $I_{4}$ defined at each vertex, a 3-index identity tensor $I_{3}$ defined at the center of each link where physical operators reside, and a $U$-matrix linking any two neighboring $I_{3}$ and $I_{4}$
\begin{equation}
  U=\frac{1}{\sqrt{2}}\left(\begin{array} {cc}
1 & 1\\
1 & -1
\end{array}\right).
\end{equation}
To find its RBM representation, we identify the bond-centered $I_{3}$ tensors as the visible units and the vertex-centered tensors $I_{4}$ as the hidden units, and decompose
$U$ as
\begin{equation}
U = \frac{1}{\sqrt{2}} \left(\begin{array}{cc}
1\\
 & e^{0}
\end{array}\right)\left(\begin{array}{cc}
1 & 1\\
1 & e^{\mathrm{i}\pi }
\end{array}\right)\left(\begin{array}{cc}
1\\
 & e^{0}
\end{array}\right).
\label{eq:U}
\end{equation}
Using \Eq{eq:AeqLWL}, we then find the connection weights and biases of visible and hidden units to be
\begin{eqnarray}
W & = & \mathrm{i}\pi, \\
a & = & 0, \label{eq:W_a_b}\\
b & = & 0.
\end{eqnarray}
The resulting wave function is depicted in Fig.~\ref{fig:Toric}(b). It is a RBM with only nearest neighboring connections between visible and hidden units. Each hidden unit couples to four visible units. Tracing out all the hidden units, the RBM becomes
\begin{equation}
\Psi_\mathrm{TC}\left(v\right) =  \prod_{+} \left( 1 + e^{i\pi\sum_{i\in +} v_i}\right). \label{eq:TC}
\end{equation}

It represents a quantum state with equal weight superpositions of closed loops where the sum of the visible variables is even on each vertex~\cite{KitaevToric}. The RBM  representation \Eq{eq:TC} is simpler than that introduced in Ref.~\cite{DLDeng2016} where the hidden units are defined on both the vertices and the plaquette centers.

The four ground states of the toric code model belong to four different topological sectors.
To differentiate these degenerate states, we first define two Wilson loop operators
\begin{equation}
X_{1} = \prod_{i\in C_{1}}\sigma^{x}_{i}, \quad
X_{2} = \prod_{i\in C_{2}}\sigma^{x}_{i},  \label{eq:X12}
\end{equation}
where $C_{1}$ and $C_{2}$ denote the paths along the horizontal and vertical directions of the lattice, indicated by the red lines in Fig.~\ref{fig:Toric}(a), respectively.
$X_{1}$ and $X_{2}$ are mutually commuting. They also commute with the Hamiltonian. Thus we can use their eigenvalues to label the eigenstates.

Both $X_{1}$ and $X_{2}$ have two eigenvalues, $+1$ and $-1$. The four-fold degenerate ground states correspond to the four eigenstates of $(X_1, X_2)$. They can be classified into four topological sectors, labeled by the eigenvalues of $(X_1, X_2)$ as $(\pm , \pm)$. The wavefunction illustrated in Fig.~\ref{fig:Toric}(b) belongs to the $(+,+)$ sector, namely $\left\langle v| \Psi^{(+,+)} \right\rangle =  \Psi_\mathrm{TC}\left(v \right)$.

Now let us introduce the following two operators along the paths $D_{1}$ and $D_{2}$ indicated by the dashed orange lines in Fig.~\ref{fig:Toric}(a),
\begin{equation}
Z_{1}  = \prod_{i\in D_{1}}\sigma^{z}_{i},  \quad
Z_{2}  =  \prod_{i\in D_{2}}\sigma^{z}_{i}, \label{eq:Z12}
\end{equation}
These two operators do not commute with $X_{1}$ and $X_{2}$. They transform the states between different topological sectors~\cite{kitaev2002classical}. For example,
$\left|\Psi^{(-,+)}\right\rangle  =  Z_{2}\left|\Psi^{(+,+)}\right\rangle$,
$\left|\Psi^{(+,-)}\right\rangle  =  Z_{1}\left|\Psi^{(+,+)}\right\rangle$, and
$\left|\Psi^{(-,-)}\right\rangle = Z_{1}Z_{2}\left|\Psi^{(+,+)}\right\rangle$.
In the PEPS representation, $Z_{1,2}$ is to change all the identity tensors $I_{3}$ shown in Fig.~\ref{fig:Toric}(a) along the $D_{1,2}$ path into diagonal tensors $\mathrm{diag}(1, -1)$. In the corresponding RBM representation, this is to change the bias of the visible variables in \Eq{eq:W_a_b} from $a=0$ to $a=\ln\left(-1\right)=\mathrm{i}\pi$ along the $D_{1}$ and $D_{2}$ paths [Fig.~\ref{fig:Toric}(c,d,e)].

\begin{figure}
\begin{centering}
\includegraphics[width=1\columnwidth]{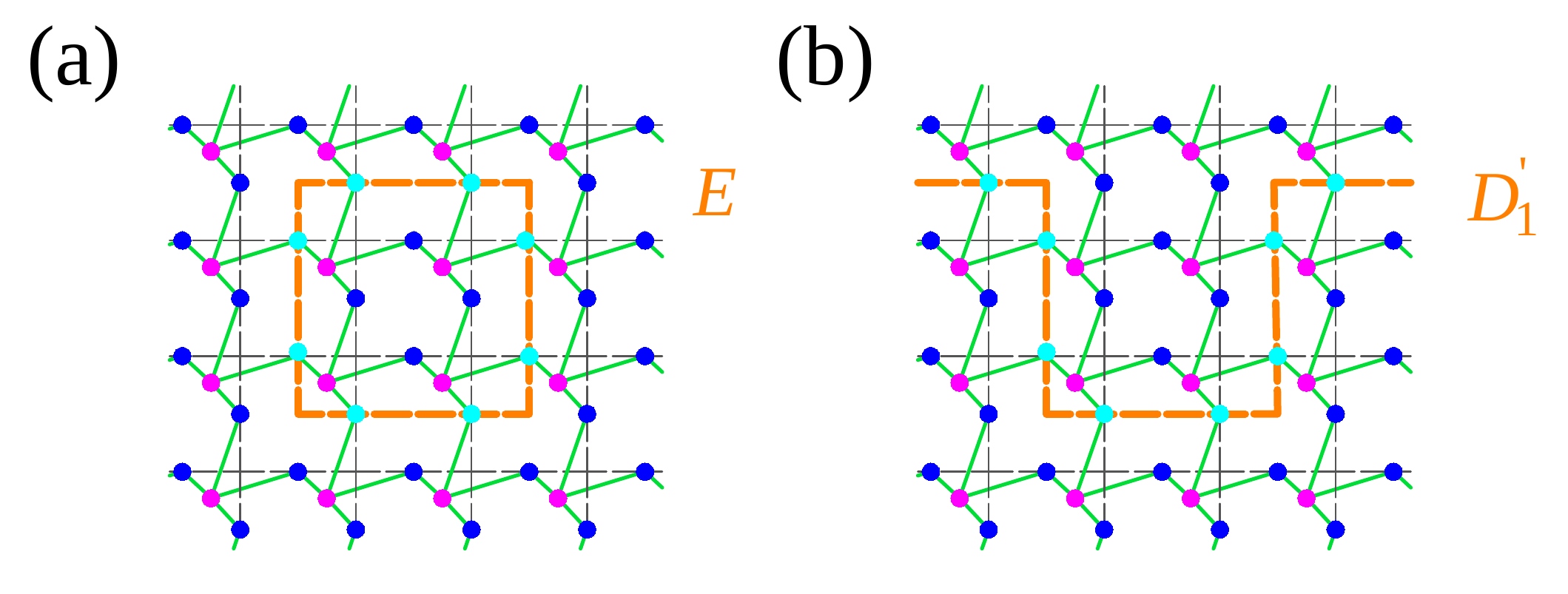}
\par\end{centering}
  \caption{Demonstration of the gauge invariance of RBM. (a) The wavefunction obtained by applying four $\mathcal{A}_{+}$ operators [\Eq{eq:AB}] within the region enclosed by the orange path $E$ to the RBM shown in Fig.~\ref{fig:Toric}(b). The visible biases are changed from 0 to $\mathrm{i}\pi$ on the path, but the two RBM wavefunctions before and after the transformation are gauge equivalent. (b) The RBM wavefunction is gauge equivalent to the that shown in Fig.~\ref{fig:Toric}(d) by moving the Wilson path from $D_{1}$  to $D'_{1}$. Please refer to Fig.~\ref{fig:Toric} for the definition of symbols.\label{fig:loop}}
\end{figure}

Note that the Wilson loops, $D_1$ and $D_2$, are not necessary to be straight lines. But they must wind around the torus (Fig.~\ref{fig:loop}).
The RBM that is obtained by continuously distorting $D_1$ or $D_2$ is gauge equivalent to the original one.  In other words, to apply the operator $X_{1}$ defined on a closed path without winding the torus to any of the ground state will not alter its topological sector. For example, the RBM shown in Fig.~\ref{fig:loop}(a) is gauge equivalent to that shown in Fig.~\ref{fig:Toric}(b). They are related by a loop of $\sigma^{z}_{i}$ operators along the path $E$ in Fig.~\ref{fig:loop}(a), which is equal to the product of all $\mathcal{A}_{+}$ operators enclosed by $E$. This product of $\mathcal{A}_{+}$ operators changes the visible biases along the closed loop $E$ from $a =0$ to $\mathrm{i}\pi$. But the wavefunction remains in the same topological sector since $\mathcal{A}_{+}$ at each vertex is conserving. Similarly, the state in Fig.~\ref{fig:loop}(b) is identical to that in Fig.~\ref{fig:Toric}(c). More detailed discussion on the redundancy of the RBM parametrization is given in Sec.~\ref{sec:redundancy}.

\section{Implications of the RBM-TNS correspondence \label{sec:application}}


\subsection{Optimizing RBM using tensor-network methods}
\label{sec:redundancy}

Similar to TNS, it is known that RBM or other neural network function contains redundant degrees of freedom~\cite{XX}. Two RBMs with different connection weights and biases may represent equivalent functions. An RBM can be simplified or optimized by removing redundant degrees of freedom using the well established tensor-network methods.

In one dimension, for example, one can use the canonicalization approach of MPS to optimize an RBM.
To do this, we first transform an RBM into an MPS using the algorithm introduced in Sec.~\ref{sec:RBM2TN}. The MPS is then canonicalized to minimize the bond dimensions for all the local tensors by discarding zero singular vectors~\cite{canonicalClassical,canonicalQuantum}. This can also partially fix the gauge of the MPS. Finally, we map this optimized MPS back to an RBM using the approach introduced in Sec.~\ref{sec:TN2RBM}. The RBM such obtained is equivalent to the original one, but is optimized.

To understand this optimization scheme, let us consider the RBM wavefunction of the 1D cluster state presented in Ref.~\cite{DLDeng2016}. This RBM contains equal number of visible and hidden units with each hidden unit connecting to three visible units. It can be mapped onto a $D=4$ MPS using the simple approach introduced in Sec.~\ref{sec:RBM2TN}
because each bipartition cuts two connections. By taking the canonical transformation, the bond dimension of this MPS is reduced to 2. Mapping this simplified MPS back, we obtain an optimized RBM with each hidden unit just connecting two neighboring visible units. See Appendix~\ref{appendix:sufficient} for detailed parametrizations of the simplified RBM.

\begin{figure}[t!]
\begin{centering}
\includegraphics[width=1\columnwidth]{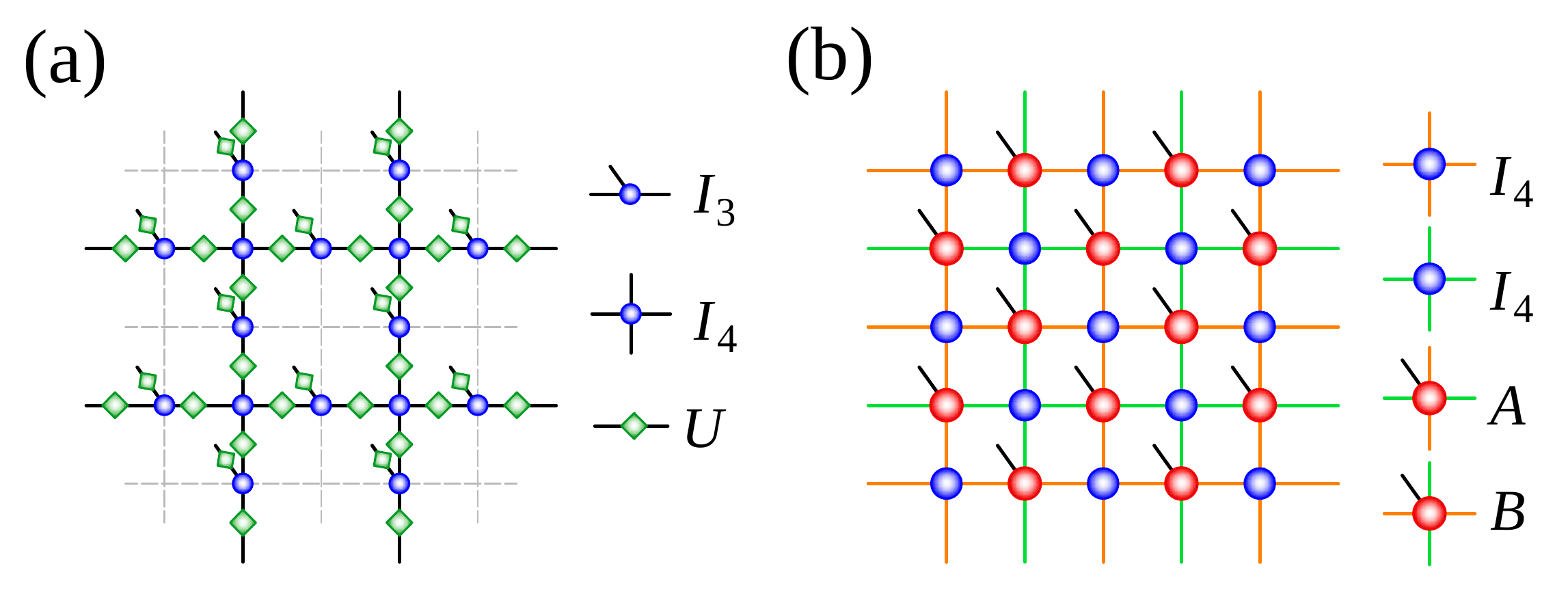}
\par\end{centering}
  \caption{Two equivalent TNS representations of the toric code ground states obtained by applying the projection operator $\mathcal{P}$ to an initial wavefunction $|\psi_0\rangle$ defined in the texts. (a) $\theta = 0$. The local identity tensor, $I_4$, is defined at the center of each plaquette, and $U$ is the matrix defined by \Eq{eq:U}. (b) $\theta$ is arbitrarily chosen. The local tensors are defined on both the original original and dual lattice sites. $A$ and $B$ are two $\theta$-dependent 5-index tensors. \label{fig:TNsForToric}}
\end{figure}

An RBM defined in higher dimensions can be also simplified by mapping it onto a PEPS or other higher-dimensional TNS. The bond degrees of freedom of the PEPS can be reduced or at least partially reduced (if there are redundancies), by taking higher-order singular value decompositions (or more generally Tucker decompositions) for all the local tensors~\cite{2012Xie}, or just singular value decompositions for all neighboring pairs of local tensors similar to that used in the determination of the ground-state wavefunction by the simple update method~\cite{2008Jiang}.

The redundancy of RBM can be exemplified using the toric code model \Eq{eq:TCH}. The ground state wavefunction can be obtained by applying the projection operator
\begin{equation}
  \mathcal{P}=\prod_{+} \frac{1+\mathcal{A}_{+}}{2} \prod_{\square}\frac{1+\mathcal{B}_{\square}}{2}
\end{equation}
to an arbitrary product state over the spins
$
 |\psi_0\rangle = \prod_i \left( \cos\theta \ket{\uparrow} + \sin\theta  \ket{\downarrow} \right)_i
 \label{eq:psi0}
$.
Apparently, the PEPS and its corresponding RBM such obtained is not unique. The state shown in Fig.~\ref{fig:Toric}(a) corresponds to the choice $\theta = \pi/4$. Figure~\ref{fig:TNsForToric}(a) shows another choice with $\theta =0$ which has $4$ index tensors in the plaquette center. It is the eigenstate of $Z_{1,2}$ in \Eq{eq:Z12}, which is the superposition of the states  Fig.~\ref{fig:Toric}(b,c,d,e). While Fig.~\ref{fig:TNsForToric}(b) shows the state obtained with a general $\theta$ which has 4 index tensors defined both in the plaquette center and on the vertices. The corresponding RBM corresponds to~\cite{DLDeng2016} and contains more connections than the other two cases. Although these are all ground states of the toric code model, it is impossible to find a local gauge transformation in the internal bonds to connect the local tensors because the ground state is a non-injective $Z_{2}$ spin liquid state~\cite{Z2injective, Z2spinliquid1,Z2spinliquid2}. Exploiting the rich math structures of non-injective PEPS~\cite{Ginjective}, one may further transform or simplify various RBM functions.

In practice, the TNS used in the simplification of an RBM may have a huge bond dimension which is difficult to handle. There are two approached that can be used to resolve this problem. The first is to dynamically truncate the TNS bond dimensions during the translation from the RBM to TNS. This can avoid the storage of huge TNS tensors. The other is to divide the system into several overlapped pieces, and perform the simplification for each piece separately. In either case, one can simplify the original RBM using the TNS canonicalization technique.


\subsection{TNS representation of the shift-invariant RBM and its entanglement capacity \label{sec:shiftinvariantRBM}}

\begin{figure}
\begin{centering}
\includegraphics[width=1\columnwidth]{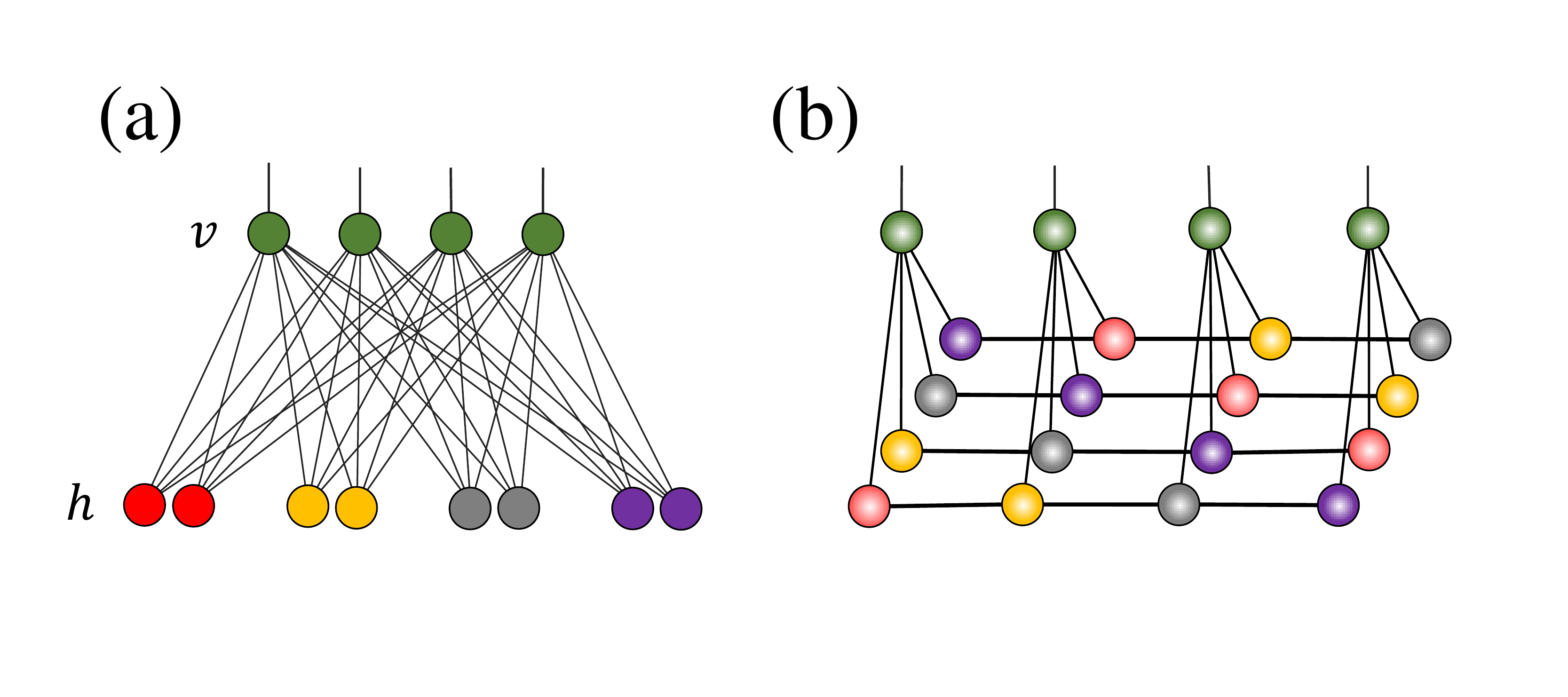}
\par\end{centering}
\caption{(a) A shift-invariant RBM of $n_v = 4$ and $n_h = 2$, which is obtained by multiplying $n_{v}$ copies of RBM with shifted connections. (b) The corresponding MPS, can be constructed by connecting $n_{v}$ copies of the MPS constructed from the corresponding RBM, each tensor offsets by one site. The green dots are identity tensors. \label{fig:shiftinvariantRBM}}
\end{figure}

The variational Monte Carlo study presented in~\cite{Carleo:2016vp} employed a shift-invariant RBM function~\cite{Anonymous:wjJtZlE6} to enforce the translational invariance of a physical system. The variational ansatz is a product of $n_{v}$ RBM functions defined in \Eq{eq:RBM3_P}
\begin{equation}
\Psi(v) = \prod_{\mathcal{T}} \Psi_\mathrm{RBM}\left(\mathcal{T} v\right),
\label{eq:siRBM}
\end{equation}
where $\mathcal{T}$ is the translational operator which shifts the visible variables around the periodic spatial direction. Figure~\ref{fig:shiftinvariantRBM}(a) shows an example of a shift-invariant RBM. Assuming each RBM wave function $\Psi_{\mathrm{RBM}}$ contains $n_{h}$ hidden units, the shift-invariant RBM contains $n_{v} n_{h}$ hidden variables.

This shift-invariant RBM can be also written as an MPS. To do this, we first express each individual factor $\Psi_{\mathrm{RBM}}$ as an MPS with bond dimension $D_\mathrm{RBM}$. For example, one of the RBMs used in Ref.~\cite{Carleo:2016vp} is a fully connected RBM with four hidden units before translational shift. When constructing the MPS, the minimum set $C$ satisfying \Eq{eq:perp} contains all the four hidden units. It corresponds to an MPS with bond dimension $D_\mathrm{RBM}=2^{4}=16$. Next, we assemble these MPSs into a tensor network as illustrated in  Fig.~\ref{fig:shiftinvariantRBM}(b). This tensor network can be further written as a single MPS. The corresponding bond dimension is $D=(D_\mathrm{RBM})^{n_{v}}$, which appears to have a much higher entanglement entropy bound than each factor $\Psi_{\mathrm{RBM}}$.

In order to better estimate the expressive power of the shift-invariant RBM, we can directly map it into an MPS using the algorithm introduced in Sec.~\ref{sec:method3}~\cite{SM}. Using this approach, one first identify the minimal interface region and determines the optimal bond dimension based on the specific structure of the shift-invariant RBM with enlarged hidden units. For example, 
the enlarged shift invariant RBM corresponds to an MPS with bond dimension $D=2^{n_{v}/2}$ in the center, see Fig.~\ref{fig:shiftinvariantRBM}(a).

The shift-invariant operation plays an important role in improving the accuracy of the variational wave function~\cite{Carleo:2016vp}
because it drastically increases the entanglement capability of the wave function without increasing the number of variational parameters. This construction can be generalized to other variational wave function \cite{peps}. Moreover, this trick can be also used to implement other symmetries, such as rotation or inversion symmetries.


From the equivalence between RBM and TNS, we find that there are several guiding principles that can be used to design even more powerful RBM variational ansatz for quantum systems. First, the connection in RBM should be global because otherwise the bond dimension of the corresponding MPS will have finite bond dimensions which do not scale with the system size. 
Note that this requirement does not mean the RBM connections have to be dense.
A sparsely connected RBM with long-range connections can also represent a TNS with large bond dimensions. In fact, the results of Ref.~\cite{Carleo:2016vp} show that many of the optimized RBM connection weights are close to zero. Reference~\cite{Mocanu2016} also shows that sparse connected RBM can have good performance in learning real dataset. Second, sharing the same parameters for many hidden and visible units can significantly reduce the number of independent variational parameters without scarifying the entanglement capacity of the ansatz. And finally, one can connect multiple hidden units to the same set of visible units of RBM to mediate large entanglement between them. 




\subsection{An entanglement perspective to unsupervised learning \label{sec:entanglement}}

A natural consequence of the connection between RBM and TNS is a quantum entanglement perspective on unsupervised learning of probabilistic models. The universal approximation theorem in machine learning~\cite{Freund:1994tu,LeRoux:2008ex,Montufar} states that there exists an RBM to describe a dataset to any accuracy, if there is no limit on the number of hidden units. By introducing the entanglement entropy of real dataset one can better quantify the required resource in terms of the number of hidden neurons and connections of an RBM, or equivalently, the effective bond dimensions of a TNS.

We first clarify the definition of entanglement entropy for real dataset. Assuming the instances of a dataset follow a probability distribution $P(v)$, we introduce a probability amplitude $\Psi(v)=\sqrt{P(v)}$ in analog to the quantum mechanical wave function. Using $\Psi(v)$, one can define the reduced density matrix and entanglement entropy of the dataset following Eqs.~(\ref{eq:Sent}, \ref{eq:rho}).
The entanglement entropy defined in this way is meaningful because it captures the complexity of real dataset similar to classical information theoretical measures~\cite{haykin2009neural}. Transferring quantum entanglement perspective to machine learning provides a practically useful way to quantify the difficulty of unsupervised learning and guides future progresses with insights in modeling quantum many-body states.
These considerations are relevant to those generative modelling inspired by quantum physics, where one uses a wavefunction square to model the probability ~\cite{pmlr-v20-bailly11,Zhao-Jaeger, Anonymous:9rpSZ_B7, 2017arXiv170901662H}. 
 
Consider a dataset of natural images, the correlations between pixels are typically dominated by short range ones which suggests that the entanglement entropy of the dataset defined in above is relative small. As a consequence, dense connections in the RBM are not absolutely necessary. In fact, the authors of Ref.~\cite{2013arXiv1312.5258D} showed that a dense RBM still performs well even when $80\%$ of the connections are randomly removed. Reference~\cite{Mocanu2016} also proposed an RBM with sparse connections with small-world network structure and found that it performs well compared to a densely connected RBM. Moreover, the distribution of the entanglement is inhomogeneous in various locations of the space. With an entanglement quantification, these features of the dataset can be exploited in the neural network structure design.

Another advantage of introducing quantum entanglement for realistic dataset is that the RBM-TNS connection may allow one to adopt the techniques developed in quantum physics directly to machine learning. For example, it is straightforward to estimate the entanglement entropy upper bound of an RBM via counting the bond dimension of its TNS representation. Alternatively, entanglement entropy is a useful characterization of the difficulty of the learning task when directly using TNS to model the dataset~\cite{Anonymous:9rpSZ_B7}.

Interestingly, the entanglement structure of deep learning was also recently explored by authors in computer science~\cite{1704.01552}. Their discussions were mainly on feedforward neural networks but the key tool for achieving their conclusions is from the techniques developed in tensor networks.

\subsection{Entanglement advantage of deep Boltzmann machines over the shallow ones \label{sec:deepBM}}
\begin{figure}[t!]
\begin{centering}
\includegraphics[width=1\columnwidth]{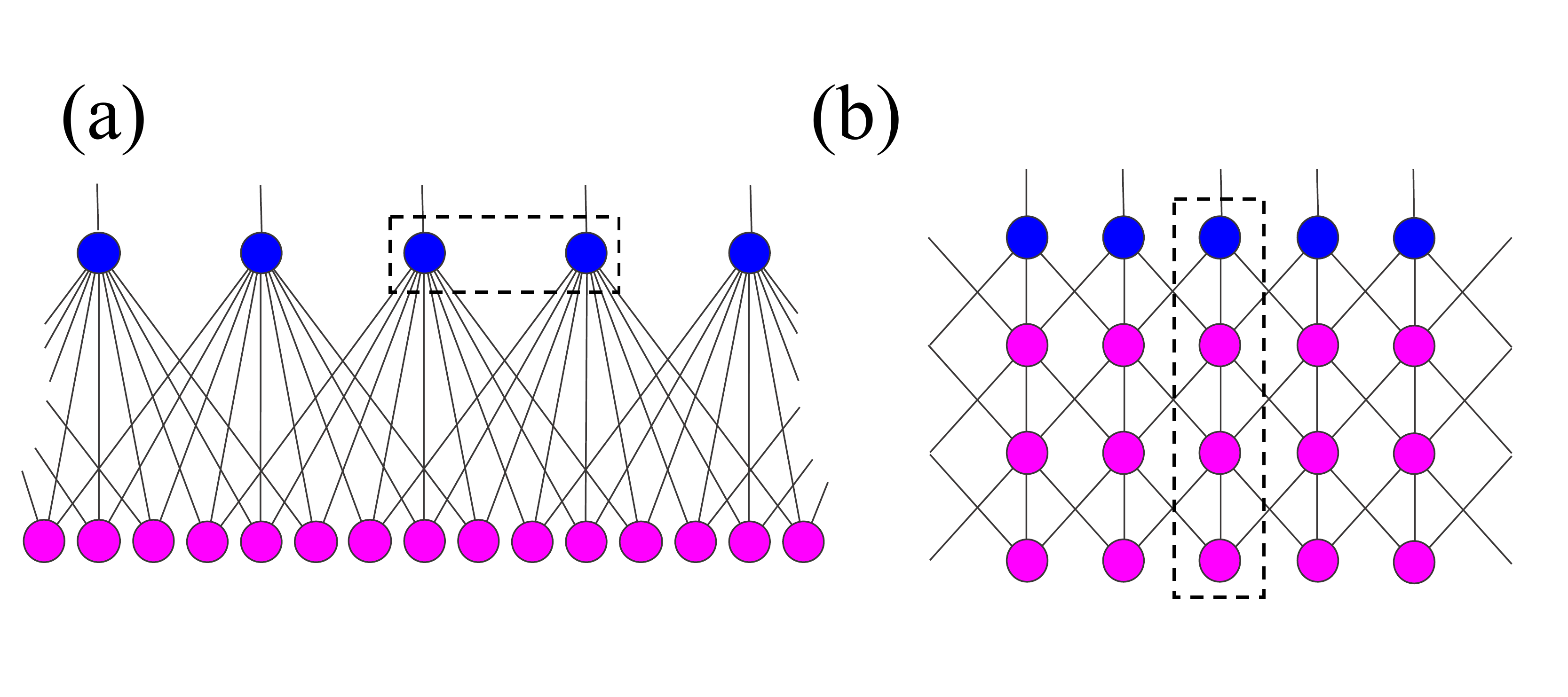}
\par\end{centering}
  \caption{(a) An RBM and (b) a deep Boltzmann Machine with the same number of $n_{v}$ visible units (blue dots), $n_{h}=3n_{v}$ hidden units (magenta dots) and $9n_{v}$ connections. According to Sec.~\ref{sec:RBM2TN} the corresponding MPS representation of the RBM and DBM had bond dimension $D_\mathrm{RBM}=2^2$ and $D_\mathrm{DBM}=2^4$ respectively. The dashed lines enclose a minimal number of units which split the network once their values are observed. The comparison shows the that DBM has larger entanglement capacity with the given number of parameters.
  \label{fig:deepRBM}}
\end{figure}

The mapping between RBM and TNS discussed in Sec.~\ref{sec:RBM2TN} is applicable general Boltzmann Machines without the bipartite graph restriction.
In particular, applying the to deep Boltzmann machines (DBM)~\cite{DBM} explains their advantage over the shallow RBMs.

The shallow and deep BM architectures shown in Fig.~\ref{fig:deepRBM}. They both have $n_{v}$ visible units, $n_{h}=3n_{v}$ hidden units and $9n_{v}$ connections. In contrast to the RBM, the hidden units of the DBM shown in Fig.~\ref{fig:deepRBM}(b) is organized into multilayer structures. Following Sec.~\ref{sec:method3} the $C$ set units are enclosed by dash lines, when they're fixed, all the visible units set $X$ is proportional to all the right visible units set $Y$. The bond dimensions are $D_\mathrm{DBM}=16$ and $D_\mathrm{RBM}=4$ respectively. Therefore, using the same amount of parameters, the DBM is able to express more complex functions with larger entanglement entropy.

As a concrete example of the application of the considerations in above, we consider the $4\times 4$ Bars and Stripes dataset~\cite{mackay2003information} used in \cite{2017arXiv170901662H} as an example. The wave function is the equal superposition of $30$ valid configurations. And the exact entanglement entropy is $\ln{15}-\frac{7}{15}\ln{7}\approx 1.80$, which is larger than $\ln{4}$. This implies that dataset can not be captured by the RBM in shown in Fig.~\ref{fig:deepRBM}(a). However, capturing the same distribution using the DBM in Fig.~\ref{fig:deepRBM}(b) is possible. 

In general, the visible units in a deep Boltzmann machine can have longer ranged effective connections  mediated by the deep hidden units (c.f. Fig.~\ref{fig:newMethod}), hence a larger entanglement capacity compared to the RBM with the same number of hidden units and connections. Applying the mapping to TNS thus offers a valuable way to analyze and compare the expressive power of Boltzmann machines with various architectures.


\section{Discussions and Outlooks \label{sec:discussion}}

We have discussed the general and constructive connection between the RBM and TNS. This equivalence sets up a bridge between the field of deep learning and quantum physics, allowing us to use the well-established entanglement theory of TNS to quantify the expressive power of RBM and obtain lower bound on the required resources compared to previously know results~\cite{Freund:1994tu,LeRoux:2008ex,Montufar}. It puts the discovered similarity between the renormalization group and deep learning~\cite{Mehta:2014ua} in a more rigorous manner, and provides a practically useful approach to remove the redundant degrees of freedom in the RBM functions (Sec.~\ref{sec:redundancy}). Moreover, connections to the TNS identify the shift-invariant construction~\cite{Anonymous:wjJtZlE6} as a key ingredient in the successful variational calculation~\cite{Carleo:2016vp} and sets up useful guiding principles to construct more powerful variational ansatz (Sec.~\ref{sec:shiftinvariantRBM}). Akin to the success of TNS in quantum physics, our finding suggests that the success of deep learning is related to the relatively low entanglement entropy in the datasets represented by the RBMs, such as natural images and speech signals (Sec.~\ref{sec:entanglement}). The entanglement entropy also offers a new perspective for using deep Boltzmann Machines (Sec.~\ref{sec:deepBM}).

The correspondence between RBM and TNS suggests that the physical insights and technical methods developed in quantum many-body physics can be exploited in the field of machine learning, and vice versa. In fact, the tensor-network methods have already been applied to the pattern recognition~\cite{Anonymous:9rpSZ_B7}. As we discussed in Sec.~\ref{sec:RBM2TN} the RBM can represent quantum state more compactly than TNS, this may offer computational advantages~\cite{Carleo:2016vp}. Deep learning algorithms~\cite{Goodfellow-et-al-2016-Book} and industrial software and hardware~\cite{nvidia, tensorflow1, tensorflow2} may also be beneficial for quantum many-body physics researches through this connection.

For further investigation, it is interesting to explore the connection between the deep learning architectures, such as the deep Boltzmann machines~\cite{DBM}, and the multi-layer TNS, such as the tree tensor networks~\cite{TTN_Murg, TTN_Silvi, TTN_Shi, TTN_Tagliacozzo} and the multi-scale entanglement renormalization ansatz~\cite{MERA}.
In passing we also note the efforts of understanding the expressive power of the deep feedforward neural networks~\cite{Cohen:2015vd, Poggio:2016uq}. We believe the insights on quantum entanglement and tensor network states can deepen our understanding on deep learning and guide better neural nets design.

\paragraph{Note added.} Recently, there appeared related works exploring representational and entanglement properties of Boltzmann Machines~\cite{2017arXiv170104844D,gao_efficient_2017,2017arXiv170106246H,kaubruegger_chiral_2017,clark_unifying_2017,glasser_neural_2017,nomura_restricted-boltzmann-machine_2017}. 

\section{Acknowledgments}
We thank Giuseppe Carleo, Dong-Ling Deng, Xiaopeng Li, Chen Fang, Xun Gao, E. Miles Stoudenmire, Hong-Hao Tu and Yi-feng Yang for helpful discussions. This work is supported by the National Natural Science Foundation of China (Grants No. 11474331, No. 11190024, and No. 11774398) and the Ministry of Science and Technology of China (Grant No. 2016YFA0302400).


\clearpage
\appendix

\section{A sufficient condition for RBM representation of MPS/PEPS and examples \label{appendix:sufficient}}

We give a sufficient condition for the MPS or PEPS to have an RBM representation. Many physically interesting thermal states and quantum wavefunction belong to this class.
For example, the toric code model discussed in Sec.~\ref{sec:ToricCode}, the statistical Ising model with external field, and the 1D/2D cluster states discussed in this appendix.

\begin{figure}
\begin{centering}
\includegraphics[width=1\columnwidth]{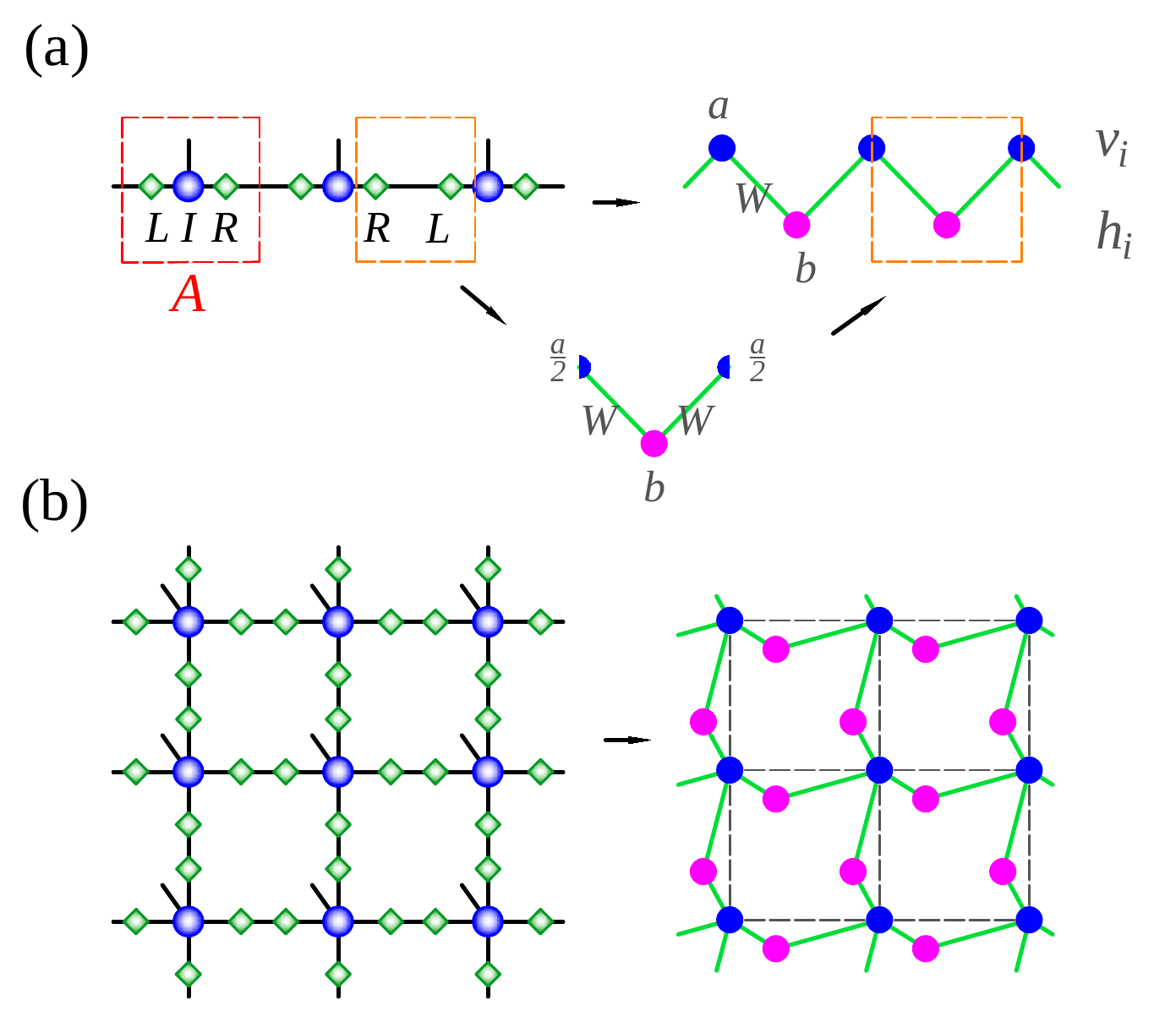}
\par\end{centering}
\caption{(a) The tensor $A$ in the dashed red box has the special form \Eq{eq:rank2mps}. The green squares denote the $L$ and $R$ matrices, while
the blue dots are the 3 index identity tensor. We transform the matrix product in the dashed orange rectangular according to \Eq{eq:Mdcomp} to obtain the RBM parameters.
The half blue dot denotes ${a}/{2}$ since the visible unit is shared by two neighboring dashed orange boxes. (b) The PEPS of a similar structure is mapped to an RBM. The meaning of the symbols is the same as (a).
\label{fig:classicalTN2RBM}}
\end{figure}

A sufficient condition for an MPS to have the RBM representation is that each tensor has the following form
\begin{equation}
A_{\alpha\beta}\left[v\right]= L_{\alpha v}R_{v\beta}, \label{eq:rank2mps}
\end{equation}
where $L$ and $R$ are two 2 by 2 matrices. As shown in Fig.~\ref{fig:classicalTN2RBM}(a), the product of $R$ and $L$ in orange dashed box can be replaced by coupling to the hidden unit of the RBM. The bias for the visible unit (blue dot) is ${a}/{2}$ because it is shared by two neighboring boxes. According to
Eqs.~(\ref{eq:Mvh}-\ref{eq:Lambdah}), we can write the $RL$ into the RBM parameters
\begin{eqnarray}
 RL &= &\left(\begin{array}{cc}
1\\
 & e^{a/2}
\end{array}\right)\left(\begin{array}{cc}
1 & 1\\
1 & e^{W}
\end{array}\right)\left(\begin{array}{cc}
1\\
 & e^{b}
\end{array}\right) \nonumber \\
 & &\left(\begin{array}{cc}
1 & 1\\
1 & e^{W}
\end{array}\right)\left(\begin{array}{cc}
1\\
 & e^{a/2}
\end{array}\right). \label{eq:Mdcomp}
\end{eqnarray}
The decomposition can be arbitrary. Here we choose a symmetric form for simplicity.

One example of this type is the statistical Ising model with the partition function,
\begin{equation}
Z = \sum_{\{s_{i}\}}\exp\left( K\sum_{\braket{i,j}}s_{i}s_{j}+H\sum_{i}s_{i}\right) \label{eq:IsingH} 
\end{equation}
where $K$ is the coupling constant and $H$ is the external field. $s_{i}=\pm 1$ are the Ising spins. To rewrite the partition function into a summation of the RBM function of the form \Eq{eq:RBM3_P} we introduce binary variables $v_{i}=(s_{i}+1)/2$.

In one dimension, the Ising partition function can be represented as an MPS shown in Fig.~\ref{fig:classicalTN2RBM}. The matrix product on each bond reads as
\begin{equation}
RL =\left(\begin{array}{cc}
e^{K+H} & e^{-K}\\
e^{-K} & e^{K-H}
\end{array}\right).\label{eq:IsingM}
\end{equation}
Combining \Eq{eq:Mdcomp} and \Eq{eq:IsingM}, we obtain the RBM parameters $W, a, b$ summarized in the first line of Table~\ref{tab:Wb}.

This procedure can be readily generalized to 2D, where the partition function is represented by a PEPS. The PEPS tensor follows a condition similar to \Eq{eq:rank2mps}, as illustrated in Fig.~\ref{fig:classicalTN2RBM}(b).
We introduce one hidden unit for each bond which couples to the two visible units connected by the lattice bond. The only difference compared to the 1D case is that the one replaces $H$ by $H/2$ in the $RL$ matrix \Eq{eq:IsingM} since the each site is shared by $4$ instead of $2$ bonds. Correspondingly, the ${a}/{2}$ bias in \Eq{eq:Mdcomp} is replaced by ${a}/{4}$ because the visible bias is also shared by $4$ connections [Fig.~\ref{fig:classicalTN2RBM}(b)]. The RBM parameters for 2D Ising model are summarized in the  second line of Table~\ref{tab:Wb}.

The above results show that a very simple sparse RBM with $n_{h}=n_{v}$ (or $n_{h}=2{n_{v}}$ in 2D) hidden units defined on the bonds can exactly reproduce the thermal distribution of the Ising model. It is remarkable that this is independent of the coupling strength and holds even at the criticality where the correlation between the visible spins are long ranged~\cite{Torlai:2016bm}. Essentially, the effect of the hidden units of the RBM play the role of a Hubbard-Stratonovich auxiliary field which decouple the interaction on the bond~\cite{1959HS}.

Another example of the TNS satisfying \Eq{eq:rank2mps} is the cluster state~\cite{ClusterState}. The MPS representation~\cite{ToricPeps, Orus2014} has $RL=\frac{1}{\sqrt{2}}\left(\begin{array}{cc}
1 & 1\\
1 & -1
\end{array}\right)$ on every bond. We can obtain the RBM parameters $W,a,b$ using \Eq{eq:Mdcomp} summarized in the Table~\ref{tab:Wb}. Similar to the Ising model case, the RBM has hidden units coupled to the physical degree of freedoms on each lattice bond. These construction is simpler than the one of Ref.~\cite{DLDeng2016} which requires each visible unit to connect to $3$ hidden units. The simplification is due to that we construct the RBM representation from the   canonical MPS of the cluster state.

\begin{table*}
\begin{centering}
\setlength{\tabcolsep}{8pt}
\renewcommand{\arraystretch}{2}
\begin{tabular}{l  c  c  c }
\toprule
Model  & $W$  & $b$  & $a$\tabularnewline
\midrule
1D Ising  & $\ln\left(4e^{4K}-2\right)$  & $-\ln\left(e^{4K}-1\right)-\ln4$  & $-4K-2H-2\ln2$  \tabularnewline
2D Ising  & $\ln\left(4e^{4K}-2\right)$  & $-\ln\left(e^{4K}-1\right)-\ln4$  & $-8K-2H-4\ln2$  \tabularnewline
1D cluster  & $\ln\frac{3}{2}$  & $-\ln2+\mathrm{i}\pi$  & $2\ln2$ \tabularnewline
2D cluster  & $\ln\frac{3}{2}$  & $-\ln2+\mathrm{i}\pi$  & $4\ln2$ \tabularnewline
\bottomrule
\end{tabular}
\par\end{centering}
\caption{The RBM parameters for the statistical Ising model
and the cluster state. Each hidden unit interacts with two visible units connected by a bond. The parametrizations are not unique and we only list one possible solution. The meaning of the Ising model parameters is given in \Eq{eq:IsingH} \label{tab:Wb}}
\end{table*}

\section{General equivalence between Boltzmann machines and TNS \label{appendix:BM2TNS}}

The name ``restricted'' in the RBM means that there are only connections between the visible and hidden units, not within them. The RBM-TNS correspondence can also be generalized to the cases without such restrictions. For example, the deep Boltzmann machines~\cite{DBM} have multilayers of hidden units with interconnections, and the Boltzmann machines (BM)~\cite{hinton1986learning} have direct connections within the visible and hidden units.

In general, the BM parametrizes a function in the form
\begin{equation}
\Psi_\mathrm{BM}\left( v\right)=\sum_{h}e^{-E\left(v, h\right)}, \label{eq:BM_p}
\end{equation}
with the energy function
\begin{equation}
E(x=v\cup h)=-\left(\sum_{i,j}W_{ij}x_{i}x_{j}+\sum_{k}\theta_{k}x_{k}\right),
\end{equation}
where $W_{ij}$ is the connection weight between the units $i$ and $j$, and $\theta_{k}$ is the bias of the unit $k$. One can either view \Eq{eq:BM_p} as a probability distribution or a complex wavefunction amplitude.



To write \Eq{eq:BM_p} into a TNS, we introduce tensor
$
M^{\left(ij\right)}=\left(\begin{array}{cc}
1 & 1\\
1 & e^{W_{ij}}
\end{array}\right)$
on the edges and diagonal tensors $\Lambda^{(k)} =\mathrm{diag}(1, e^{\theta_{k}}) $ on the vertices. 
For the visible units, the diagonal tensors have an additional dimension corresponding to the external degree of freedoms. Using these tensors, the BM \Eq{eq:BM_p} can be written as a tensor network state
\begin{equation}
\Psi_\mathrm{TNS}\left(v\right)=\mathrm{Tr}\left(\prod_{ i,j }M^{\left(ij\right)}\prod_{k}\Lambda^{\left(k\right)}\right).
\end{equation}


Conversely, one can also attempt to map a TNS back to a BM. First, we prove that any TNS constructed rank 2 tensors only
can be directly mapped to a binary BM. The CP decomposition or rank decomposition~\cite{ranktensor} of $d_{1}\times d_{2}\times\cdots\times d_{n}$
tensor $T^{(i)}$ reads

\begin{equation}
T^{(i)}_{\alpha_{1}\alpha_{2}\cdots \alpha_{n}}=\sum_{k=1}^{r}P_{\alpha_{1}k}^{\left(i\right)}Q_{\alpha_{2}k}^{\left(i\right)}R_{\alpha_{3}k}^{\left(i\right)}\cdots ,\label{eq:Tdecomp}
\end{equation}
where $P^{\left(i\right)},Q^{\left(i\right)},R^{\left(i\right)},\cdots$
are matrices. The equation holds for a minimal number $r$, which is the rank of the tensor.
For example, $r=1$ if $T$ is a vector and $r$ equals
the smaller dimension if $T$ is a matrix. While for $n\geqslant3$, $r$
is not necessarily smaller than any $d_{i}$.

A TNS can be mapped to a BM with binary units if $r=2$ and  $d_{i}=2, \forall i$. When contracting the first index of $T^{\left(i\right)}$ and the second index of $T^{\left(j\right)}$,
we can perform rank decomposition and obtain a two by two matrix on the connection
\begin{eqnarray}
M^{\left(ij\right)} & = & \left(P^{\left(i\right)}\right)^T Q^{\left(j\right)}=\left(\begin{array}{cc}
p & q\\
r & s
\end{array}\right),\label{eq:M2}
\end{eqnarray}
where $P^{(i)}$ ($Q^{(j)}$) is the matrix obtained from the decomposition \Eq{eq:Tdecomp} of $T^{\left(i\right)}$($T^{\left(j\right)}$)
respectively.  With the same procedure in Eqs.~(\ref{eq:AeqLWL},\ref{eq:W}),
we find that $W_{ij}=\ln\frac{ps}{qr}$. The two diagonal matrix
in \Eq{eq:AeqLWL} can be absorbed into the diagonal tensor,
which contributes to $\theta_{i}$. 

For more general TNS with larger ranks, we can always perform the CP decomposition following the same procedure and resulting BM will have hidden units with multistates.



\bibliography{RBMandTN}

\end{document}